\documentclass[conference]{IEEEtran}
\IEEEoverridecommandlockouts

\usepackage{cite}
\usepackage{amsmath,amssymb,amsfonts}
\usepackage{graphicx}
\usepackage{booktabs}
\usepackage{multirow}
\usepackage{xcolor}
\usepackage{tikz}
\usetikzlibrary{positioning,arrows.meta,fit,calc,backgrounds}
\usepackage[hidelinks]{hyperref}
\usepackage{microtype}
\usepackage{siunitx}
\usepackage{graphicx}
\usepackage{ulem}
\usepackage{hyperref}
\usepackage{comment}
\usepackage{balance}
\usepackage{float}
\usepackage{cuted}
\usepackage{capt-of}   
\usepackage{tabularx}

\graphicspath{{figs/}}
\newcommand{\E}{\mathbb{E}}

\begin{document}

\title{

Stateless Stochastic Computing for Generative AI on Superconducting Logic:\\A Model--Hardware Co-Design
}

\author{\IEEEauthorblockN{Mehdi Kamal and Massoud Pedram}
\IEEEauthorblockA{Ming Hsieh Department of Electrical and Computer Engineering, University of Southern California\\
\{mehdi.kamal, pedram\}@usc.edu}}

\maketitle

\begin{abstract}
Thermodynamic computing lets stochastic physical dynamics carry out a computation and
reads out the result as a time average. Recently, thermodynamic computing has been demonstrated on several hardware platforms, including digital CMOS circuits based on stochastic computing.
Superconducting single-flux-quantum (SFQ) logic
offers very high clock rates and very low switching energy, but it heavily penalizes
state, memory, fan-out, and feedback, i.e., exactly the per-neuron accumulators on which
such designs rely.
We therefore realize thermodynamic computing in SFQ logic through a co-design of a
generative model and its circuit in which the datapath holds no state at all. The model
is a class-conditional flow-matching image generator, which transforms Gaussian noise
into an image of the requested class along a short, deterministic path of four Euler
steps; in hardware, these steps are unrolled into a feed-forward pipeline of stochastic
bitstream operators, whose output is read out as a time average, as in thermodynamic
computing, without any relaxation phase.
Each layer of the model is a
butterfly-factorized linear map built exclusively from multiplexers that operate on
split-unipolar bitstreams. The ReLU activation, which in stochastic computing requires a
per-neuron up/down counter (accounting for 60--70\% of the JJs in a counter-based design),
is replaced by a stateless AND-NOT gate whose one-clock input
skew is provided at no cost by SFQ gate-level pipelining. The models are trained directly
on the stream representation of the hardware, with a regularizer that bounds the stream
gains. Synthesis with an SFQ design flow shows that stochastic streams tolerate relaxed
path balancing, which reduces the Josephson-junction (JJ) count of a hard-wired stage by 40--47\%. On
the digits, Fashion-MNIST, and grayscale CIFAR-10 datasets, the stateless design matches
the sample quality of the ReLU at equal readout noise to within about one FID point, while
requiring $1.6$--$1.9\times$ fewer JJs than a ReLU with a 4-bit counter and
$2.5$--$3.1\times$ fewer than one with an 8-bit counter (e.g., 2.8\,M instead of 5.4\,M
and 8.8\,M JJs for a four-step flow chain with hard-wired weights).
Compared to an equivalent bit-parallel binary fixed-point SFQ accelerator, the stateless stochastic
design is 32--51$\times$ smaller and reduces the required off-chip memory bandwidth by
$\sim 1000\times$ ($21\text{ Gb/s}$ vs. $28\text{ Tb/s}$), at the cost of a longer time per sample,
bringing cryogenic generative AI within realistic thermal budgets.
\end{abstract}


\section{Introduction}
\label{sec:intro}

Generative models now account for a large and growing share of AI compute. Whether they
produce images by iterative denoising~\cite{ho2020ddpm,song2021sde} or by learned
transport~\cite{lipman2023flow,liu2023rectified}, most of their energy is spent on long
sequences of dense linear maps interleaved with pointwise nonlinearities, executed on
conventional CMOS accelerators.
\emph{Thermodynamic computing}~\cite{conte2019thermodynamic,coles2023thermodynamic}
proposes a different relationship with the hardware: instead of computing exact values and
then injecting pseudo-random noise, it lets stochastic dynamics carry out the computation
and obtains the result as a time average. A recent digital thermodynamic computer,
CN101~\cite{holdijk2026cn101}, realizes this idea in standard CMOS with stochastic
computing: each layer of a neural network is computed as the stationary expectation of
accumulator-based stochastic dynamics, the layers of a deep model relax concurrently
rather than in sequence, and the precision of the result is set by the length of the
averaging window. CN101 has been demonstrated on conditional VAEs and on flow-matching
models.

Computing by relaxation, however, has a price. The result is correct only once the
dynamics have settled: until then, the readout carries a bias from the initial state,
and the end-to-end latency is set by the time the joint state of all layers needs to
relax. In addition, the dynamics are carried by per-neuron state---multi-bit
accumulators and per-neuron pseudo-random generators---which is inexpensive in CMOS but,
as discussed below, costly in superconducting logic. Modern generative models, in turn,
are increasingly formulated as \emph{non-equilibrium} processes. Diffusion models
originated as a non-equilibrium thermodynamic process~\cite{sohldickstein2015nonequilibrium},
and flow matching and rectified flow~\cite{lipman2023flow,liu2023rectified} learn a
deterministic transport from noise to data that can be integrated in a few steps. We ask
whether such a model can be realized in stochastic logic by a datapath that requires no
relaxation and holds no state.

The common approach to accelerating such workloads is to design dedicated CMOS
accelerators. Emerging technologies offer an alternative, and superconducting
single-flux-quantum (SFQ) logic~\cite{likharev1991rsfq} is among the most mature of them.
SFQ gates switch within picoseconds and dissipate attojoule-scale energy per switching
event, which enables very high clock rates at very low power. These advantages, however,
come with constraints that make SFQ poorly suited to the conventional building blocks of
accelerators. First, every gate is clocked, so logic depth becomes pipeline depth and
unequal paths must be balanced with flip-flops. Second, the fan-out is one, so every branch
requires a splitter. Third, memory is scarce and expensive. Fourth, a feedback loop can
advance only once per loop latency. Wide adders, multipliers, and large on-chip memories
are therefore costly in SFQ.

This paper targets the implementation of a generative AI model on superconducting
circuits. We propose a model--hardware co-design that realizes flow-matching generation
in stochastic logic with a stateless datapath, while respecting the limitations of SFQ
technology. On the model side, we unroll a four-step flow-matching sampler into a
feed-forward pipeline of stochastic bitstream operators. Because the pipeline contains
no per-neuron state and no feedback, every register bit is a fixed function of a finite
set of random draws, so the stream statistics are exact from the first cycle after the
pipeline has filled, without a relaxation phase. On the hardware side, stochastic computing (SC) suits SFQ well: the
picosecond gates make the long bitstreams of SC affordable in latency, and an SC network
fits the first three constraints as long as it avoids adders and wide state. A naive SC
neural network, however, still needs per-neuron counters to implement its nonlinearities.
In our initial design, these counters alone accounted for 60--70\% of the Josephson
junctions (JJs) and introduced an eight-level feedback loop in every neuron. We remove them
by co-designing the activation function, the training method, and the circuit. 

The main contributions of this paper are as follows:
\begin{itemize}
\item A \textbf{stateless generative pipeline in stochastic logic}: class-conditional
flow matching with four unrolled Euler steps, each realized by a single multiplexer per
pixel, whose stream statistics need no relaxation phase and whose residual stages
increase the stream gain by at most a bounded factor per step (Sec.~\ref{sec:model}).

\item An \textbf{adder-free network form with stationary weights}: butterfly-factorized
layers~\cite{dao2019butterfly} are rewritten exactly into multiplexers, sign routing, and
host-side gains, with a fan-out of two and a single random-number generator per stage. The
weights become 8-bit select thresholds that are written once per model, and the weight
memory is $7$--$37\times$ smaller than that of dense layers of the same width
(Sec.~\ref{sec:model-constraints},~\ref{sec:model-nf}).

\item A \textbf{stateless activation function}: an AND-NOT gate on the two channels of a
split-unipolar stream, $a(t)\wedge\neg b(t-1)$, replaces the ReLU counter. The required
one-clock skew comes at zero hardware cost 
from the inverter in an SFQ gate-level pipeline, so the activation
costs one inverter and one AND gate per neuron and has no feedback loop
or per-neuron accumulator state
(Sec.~\ref{sec:model-act},~\ref{sec:hw-act}).

\item A \textbf{stream-space training method}: the network is trained on the expectation of
the hardware under an explicit independence model, including the common mode of the
streams seen by the AND-NOT activation, with an unrolled-chain refinement and a tail
regularizer that bounds pixel variance and eliminates readout noise divergence
on the stream gains (Sec.~\ref{sec:model-train}).

\item A \textbf{formal and empirical treatment of operand independence}: we prove that the
MUX means are exact under ideal random-number generators, show that layer-specific
power-of-two delays make the AND-NOT activation exact in every layer, and measure the
residual covariances of the implemented generators in long bitstream simulations
(Sec.~\ref{sec:model-indep},~\ref{sec:res-indep}).

\item An \textbf{SFQ accelerator architecture and cost model}, calibrated against
qPALACE~\cite{qpalace} logic synthesis, showing that stochastic streams tolerate relaxed
path balancing, which saves 40--47\% of the JJs per stage by omitting path-balancing flip-flops on data inputs (Sec.~\ref{sec:hw}).

\item A \textbf{verification ladder} from the floating-point model to the RTL, which is
bit-exact against a golden cycle-accurate simulator, and an evaluation on three datasets
showing ReLU-level generative quality with $1.6$--$3.1\times$ fewer JJs, including a
comparison with a binary fixed-point SFQ implementation of the same networks
showing a 32--51$\times$ area reduction and a $\sim 1000\times$ off-chip 
memory bandwidth reduction ($21\text{ Gb/s}$ vs. $28\text{ Tb/s}$)
(Sec.~\ref{sec:results}).
\end{itemize}

The remainder of the paper is organized as follows. Section~\ref{sec:background} reviews
stochastic computing, the generative models used in this work, and SFQ logic.
Section~\ref{sec:model} presents the model design, from the generative formulation to the
activation function and training method. Section~\ref{sec:hw} describes the SFQ
accelerator from the system level down to the cell level, together with the cost model.
Section~\ref{sec:results} reports and discusses the experimental results, and
Section~\ref{sec:conclusion} concludes the paper.

\section{Background and Related Work}
\label{sec:background}

\subsection{Stochastic Computing}
\label{sec:bg-sc}

In stochastic computing (SC)~\cite{gaines1969stochastic,alaghi2013survey}, a number
$p\in[0,1]$ is represented by a bitstream $z(t)$ whose bits are 1 with probability $p$.
A stochastic number generator (SNG) produces such a stream by comparing a pseudo-random
number $R(t)$ with a threshold $X=p\,2^B$, i.e., $z(t)=[R(t)<X]$. The value is recovered by
counting, $\hat p = \frac{1}{T}\sum_t z(t)$, with a standard deviation of
$\sqrt{p(1-p)/T}$ for independent bits. Arithmetic then reduces to simple logic. For
example, a two-input multiplexer (MUX) whose select stream has probability $s$ produces an
output stream with mean
\begin{equation}
\E[z] = s\,\E[u] + (1-s)\,\E[w],
\label{eq:mux}
\end{equation}
i.e., a scaled addition. Eq.~\eqref{eq:mux} holds whenever the select is
\emph{independent} of the data inputs; the data inputs themselves may be arbitrarily
correlated. We exploit this asymmetry---independence is needed only between select and
data---throughout our designs.

Signed values require a dedicated representation. We use a \emph{split-unipolar} pair of
streams $(z^+,z^-)$ that represents $\E[z^+]-\E[z^-]$, together with a positive real
\emph{gain} $g$ kept outside the hardware, so that the represented value is
$y=g\,(\E[z^+]-\E[z^-])$. Nonlinear functions in SC are conventionally implemented with
finite-state machines driven by the stream~\cite{brown2001stochastic}; in particular, a
ReLU on a split-unipolar pair requires an up/down counter (Sec.~\ref{sec:model-act}).

\subsection{Generative Models}
\label{sec:bg-gen}

\paragraph*{Variational autoencoders} A conditional VAE~\cite{kingma2014vae,sohn2015cvae}
trains an encoder $q(z\,|\,x,c)$ and a decoder $p(x\,|\,z,c)$ by maximizing the evidence
lower bound (ELBO). Samples of class $c$ are generated by decoding $z\sim\mathcal{N}(0,I)$.
Since only the decoder runs at inference time, the inference workload is a single
feed-forward network.

\paragraph*{Diffusion and flow matching} Diffusion
models~\cite{sohldickstein2015nonequilibrium,ho2020ddpm,song2021sde} learn to reverse a
noising process; their sampler is a discretized stochastic or ordinary differential
equation that requires tens to thousands of network evaluations. Flow
matching~\cite{lipman2023flow} and rectified flow~\cite{liu2023rectified} instead regress a
velocity field $v(x,t)$ that transports the noise distribution at $t=0$ to the data
distribution at $t=1$ along the linear interpolation $x_t=(1-t)x_0+t\,x_1$:
\begin{equation}
\mathcal{L}_{\mathrm{FM}} = \E_{t,x_0,x_1}\,\big\| v_\theta(x_t,t,c) - (x_1-x_0)\big\|^2 .
\end{equation}
The \emph{reflow} procedure~\cite{liu2023rectified} retrains the model on pairs
$(x_0,\hat x_1)$ generated by the model itself. This straightens the transport paths so
that very few Euler steps suffice, which is what makes a short, unrolled, non-equilibrium
sampler possible.

\subsection{Superconducting SFQ Logic}
\label{sec:bg-sfq}

Rapid single-flux-quantum (RSFQ) logic~\cite{likharev1991rsfq} encodes a bit as the
presence or absence of a voltage pulse carrying one flux quantum within a clock period.
Gates are built from Josephson junctions (JJs), and almost all of them are clocked: each
gate stores the pulses that arrive within a clock period and emits its result at the next
clock edge. A multi-level combinational function therefore becomes a gate-level pipeline,
and a gate whose inputs arrive from different logic levels combines data from different
waves unless the shorter paths are padded with D flip-flops (DFFs), a process called
\emph{path balancing}~\cite{pasandi2019pbmap}. In addition, SFQ gates have a fan-out of
one, so a net with $r$ consumers needs $r-1$ splitters, and the clock itself is distributed
through a splitter tree. SFQ synthesis and technology-mapping tools such as
qPALACE~\cite{qpalace} insert balancing flip-flops and splitters automatically. We report
JJ counts, the standard area metric of this technology, for the cell library of the MIT
Lincoln Laboratory process~\cite{tolpygo2016process}.

\subsection{Related Work}
\label{sec:related}

\paragraph*{Digital SFQ neural-network accelerators} Several works map conventional deep
neural networks onto SFQ logic. SuperNPU~\cite{ishida2020supernpu} introduced a
systolic-array neural processing unit built from SFQ circuits, and
SMART~\cite{zokaee2021smart} showed that the shift-register-based scratchpad memories of
such accelerators limit them to less than 40\% of their peak throughput, highlighting the
memory bottleneck of the technology. JBNN~\cite{fu2022jbnn} implemented a binarized neural
network in SFQ and observed that, because every SFQ gate is a pipeline stage, conventional
accumulator designs are unsuitable. To relax area constraints,
U-SFQ~\cite{gonzalezguerrero2022usfq} encodes values in pulse streams and race logic,
reducing the JJ count by up to $200\times$ at the cost of latency. All of these designs
target discriminative inference and rely on explicit accumulation.

\paragraph*{AQFP-based neural networks} Adiabatic quantum-flux-parametron (AQFP) logic is
the most energy-efficient superconducting logic family. Cai et~al.~\cite{cai2019aqfpsc}
developed the first SC-based DNN framework on AQFP and noted that its deep pipelining makes
accumulators and counters hard to realize, which motivated neurons based on majority
chains and sorting networks. SupeRBNN~\cite{li2023superbnn} exploited the randomized
behavior of AQFP devices in a randomized binary neural network through software--hardware
co-optimization. Both works address image classification rather than generation, and
target AQFP rather than RSFQ logic.

\paragraph*{Superconducting neuromorphic hardware} The pulse-based nature of SFQ logic
closely resembles neuronal spiking, which has motivated a large body of neuromorphic
work, ranging from Josephson-junction neuron models~\cite{crotty2010jjneurons} and
synaptic weighting circuits~\cite{schneider2020synaptic} to complete spiking neural network
(SNN) architectures~\cite{bozbey2020sfqsnn,karamuftuoglu2023jjsoma}. An unsupervised
SFQ-based SNN with dynamically regulated neuron thresholds was demonstrated
in~\cite{karamuftuoglu2024unsupervised}, and a high-fan-in differential neuron with
ternary synaptic connections, in which positive and negative inductive coupling realize
excitatory and inhibitory inputs, reached 96.1\% accuracy on
MNIST~\cite{karamuftuoglu2025ternary}. This differential structure is conceptually close
to the split-unipolar representation used here. SUSHI~\cite{liu2023sushi} reported a
fabricated superconducting neuromorphic chip with about $10^5$ JJs.

\paragraph*{Stochastic computing on superconducting logic} Early work proposed an SFQ
up/down counter for neural computation with stochastic logic~\cite{onomi2008updown}---the
very per-neuron counter that our stateless activation eliminates. More recently, MUX-based
stochastic adders driven by superconducting random-number generators have been studied,
and a deterministic stochastic adder was proposed to reduce the area and error of the SFQ
MUX~\cite{han2026stochadder}. Edwards et~al.~\cite{edwards2024stochastic} exploited
thermal noise to build stochastic SFQ synapses and demonstrated a rate-coded multi-layer
network on the XOR task, suggesting that physical noise sources could replace
pseudo-random generators.

\paragraph*{Stochastic and thermodynamic hardware for generative AI}
CN101~\cite{holdijk2026cn101} is a digital thermodynamic computer in CMOS that evaluates
each layer of a neural network as the stationary expectation of stochastic dynamics,
using split-unipolar bitstreams and AND/OR stochastic matrix multipliers. The multi-bit
result of each multiplier is passed to an accumulator-based neuron bank, every neuron
owns a 65-bit LFSR, and the layers of a model relax concurrently toward the result.
CN101 has run a conditional VAE on a single chip and flow-matching models across
multiple chips. Our design shares the stochastic representation but targets SFQ logic,
in which per-neuron accumulators and generators are expensive: it replaces accumulation
by multiplexer trees, the activation by a stateless gate, and per-neuron generators by
one generator per stage, so that the datapath has no state and needs no relaxation.

\paragraph*{Summary} All of the superconducting works above target discriminative tasks
such as classification, or individual arithmetic and neuromorphic primitives, and the
stochastic generative hardware of~\cite{holdijk2026cn101} relies on per-neuron
accumulator state. To the best of our knowledge, no prior work implements a generative
AI model on superconducting circuits, and none realizes an SC neural network whose
neurons are entirely stateless. This work addresses both gaps.

\section{Model Exploration and Design}
\label{sec:model}

This section presents the model top-down. We first state the hardware constraints that
shape the model and then introduce the generative formulation and the network
architecture. We then describe the exact transformation of the network into stochastic
operators, the proposed activation function, and the training procedure.
Fig.~\ref{fig:model} summarizes the main elements.

\begin{figure*}[t]
\centering
\resizebox{\textwidth}{!}{%
\begin{tikzpicture}[
  font=\footnotesize, >={Stealth[length=1.6mm]},
  blk/.style={draw, rounded corners=2pt, align=center, minimum height=11mm, fill=white,
              inner sep=2.5pt},
  lay/.style={blk, fill=blue!7, minimum width=17mm},
  act/.style={blk, fill=orange!14, minimum width=13mm},
  mux/.style={blk, fill=green!10},
  aux/.style={blk, fill=gray!10, minimum height=8mm},
  stepbox/.style={draw, rounded corners=2pt, fill=blue!7, minimum width=15mm, minimum height=7mm},
  eq/.style={draw, rounded corners=2pt, fill=gray!5, align=left, inner sep=4pt},
  ttl/.style={font=\footnotesize\bfseries, anchor=west},
  node distance=4.5mm]

\node[ttl] at (-1.2, 5.3) {(a) Unrolled non-equilibrium transport ($N=4$ Euler steps, $t_k=k/N$)};
\node (x0) at (0.2, 4.4) {$x_0\sim\mathcal{N}(0,I)$, class $c$};
\node[stepbox, right=6mm of x0] (k0) {step 0};
\node[stepbox, right=5mm of k0] (k1) {step 1};
\node[stepbox, right=5mm of k1] (k2) {step 2};
\node[stepbox, right=5mm of k2] (k3) {step 3};
\node[right=6mm of k3] (xN) {$x_4$ = sample};
\draw[->] (x0) -- (k0); \draw[->] (k0) -- (k1); \draw[->] (k1) -- (k2);
\draw[->] (k2) -- (k3); \draw[->] (k3) -- (xN);
\node[right=8mm of xN, align=left, font=\scriptsize] {each step: its own copy of the network\\
(per-step tables, own RNGs); weights $\theta$ shared in training};

\node[ttl] at (-1.2, 3.25) {(b) One Euler step $k$: $x_{k+1}=\alpha_k x_k+\beta_k f_\theta(x_k,t_k,c)$};
\node (xk) at (-0.6, 0) {$x_k$};
\node[blk, right=5mm of xk] (lanes) {lane assembly\\$[\,x_k \mid \mathbf{1}_c^{\times 4} \mid \mathbf{0}\,]$\\$n$ lanes};
\node[lay, right=of lanes] (b1) {butterfly\\layer 1\\+ bias};
\node[act, right=of b1] (a1) {AND-NOT};
\node[lay, right=of a1] (b2) {butterfly\\layer 2\\+ bias};
\node[act, right=of b2] (a2) {AND-NOT};
\node[lay, right=of a2] (b3) {butterfly\\layer 3\\+ bias};
\node[act, right=of b3] (a3) {AND-NOT};
\node[lay, right=of a3] (b4) {butterfly\\layer 4\\+ bias\\(linear)};
\node[blk, right=of b4] (fk) {$f_\theta$:\\first $P$\\lanes};
\node[mux, right=6mm of fk] (res) {residual MUX\\$\alpha_k x_k + \beta_k f_\theta$};
\node[right=5mm of res] (xk1) {$x_{k+1}$};
\foreach \u/\v in {xk/lanes, lanes/b1, b1/a1, a1/b2, b2/a2, a2/b3, b3/a3, a3/b4, b4/fk, fk/res, res/xk1}
  \draw[->] (\u) -- (\v);
\draw[->] ($(xk.east)+(1.5mm,0)$) -- ++(0,-10mm) -| node[pos=0.38, below, font=\scriptsize]
  {skip: data inputs need no alignment (no delay matching)} (res.south);
\node[aux, above=7mm of lanes] (tk) {$t_k$ $\rightarrow$ Fourier features $\phi(t_k)$};
\draw[->] (tk.east) -| node[pos=0.75, right, font=\scriptsize] {per-step first-layer bias $W_t\,\phi(t_k)$} (b1.north);
\draw[->] (tk.north) -- ++(0,3.5mm) -| node[pos=0.15, above, font=\scriptsize]
  {learned skip $\kappa(t_k)=\sigma(W\phi(t_k))$: $\alpha_k=1-a_k+a_k \kappa(t_k)$, $\beta_k=a_k=\frac{\Delta t}{1-t_k}$} (res.north);
\node[anchor=north west, font=\scriptsize, align=left] at ($(res.south east)+(1mm,-1mm)$) {$\alpha_k,\beta_k\ge0$: one MUX per pixel,\\gain $g'=\alpha_k g_x+\beta_k g_f$\\$\le(1{+}a_k\kappa)\max(g_x,g_f)$};

\node[ttl] at (-1.2, -2.55) {(c) Inside a layer, and how it is trained};
\node (xi) at (0.0, -3.5) {$x_i$};
\node (xj) at (0.0, -4.7) {$x_j$};
\node (yi) at (2.6, -3.5) {$y_i$};
\node (yj) at (2.6, -4.7) {$y_j$};
\draw[->] (xi) -- node[above, font=\scriptsize] {$a$} (yi);
\draw[->] (xj) -- node[below, font=\scriptsize] {$d$} (yj);
\draw[->] (xj) -- node[pos=0.28, left, font=\scriptsize] {$b$} (yi);
\draw[->] (xi) -- node[pos=0.28, left, font=\scriptsize] {$c$} (yj);
\node[font=\scriptsize, align=center] at (1.3, -5.45) {stage $k$: $j=i\oplus 2^k$;\\$m=\log_2 n$ stages per layer};
\node[eq, anchor=north west] (nf) at (3.4, -3.0) {%
\textbf{stochastic normal form} (exact)\\[1pt]
$a'=w_{\mathrm{self}}\,g_i,\;\; b'=w_{\mathrm{partner}}\,g_j$\\
$G_i=|a'|+|b'|,\;\; s_i=|a'|/G_i$\\
$z_i^{\pm}=\mathrm{MUX}\big(R_k(t)<S_i;\;\mathrm{route}(x_i),\mathrm{route}(x_j)\big)$\\
sign routing = channel swap; bias = MUX with 0/1 stream};
\node[eq, anchor=north west, fill=orange!8] (an) at ($(nf.north east)+(4mm,0)$) {%
\textbf{stateless activation}\\[1pt]
$\mathrm{out}(t)=z^{+}(t)\wedge\neg z^{-}(t-d_\ell)$\\
$\E[\mathrm{out}]=p^{+}(1-p^{-})$, output $(\mathrm{out},0)$\\
delay $d_\ell=1$, or $1,2,4,\dots$ (exact in every layer)\\
replaces the ReLU up/down counter:\\
no state, no feedback loop};
\node[eq, anchor=north west, fill=blue!4] (tr) at ($(an.north east)+(4mm,0)$) {%
\textbf{training in stream space}\\[1pt]
$Y\!\leftarrow\! W_kY,\; C\!\leftarrow\!|W_k|C,\; g\!\leftarrow\!|W_k|g\quad (C=g(p^++p^-))$\\
$p^{\pm}=(C\pm Y)/2g$ $\rightarrow$ activation on $p^{\pm}$\\
teacher-forced $C$ (phase 1), unrolled-chain reflow\\
loss $+\,\lambda\,\big[\,\mathrm{mean}\log(1{+}g)+\mathrm{softmax}_\tau\log(1{+}g)\,\big]$};
\draw[densely dotted] (b2.south) -- ++(0,-12mm) -| (nf.north);
\draw[densely dotted] (a2.south) -- ++(0,-14mm) -| (an.north);
\end{tikzpicture}}
\caption{Overview of the proposed model. (a)~A sample is generated by a finite-time,
non-equilibrium transport from noise to data, realized as four unrolled Euler steps.
(b)~Each step consists of four butterfly layers with stateless AND-NOT activations,
conditioned on the class (through the label lanes) and on time (through a per-step
first-layer bias), followed by a residual MUX that implements the Euler update with
non-negative weights. (c)~Each butterfly stage is rewritten exactly into multiplexers
operating on split-unipolar streams with host-side gains; the activation operates on the
two channel probabilities, and the network is trained on the expected stream statistics
($Y$, $C$, $g$) of the hardware.}
\label{fig:model}
\end{figure*}

\subsection{Design Constraints}
\label{sec:model-constraints}

The SFQ target imposes four constraints on the model: (i)~the datapath may contain no
adders or multipliers, only gates operating on bitstreams; (ii)~there may be no
per-neuron state and no feedback loops; (iii)~each stream may feed at most two consumers,
i.e., one splitter; and (iv)~since memory is scarce in SFQ, the weights must be few and
stationary, i.e., written once per model and only read during inference. All operations
that require general-purpose arithmetic---encoding the input, applying the gains, and
subtracting and scaling the readout counts---run on the host once per sample rather than
once per clock cycle.

Regarding (iv), each weight is read in every clock cycle (Sec.~\ref{sec:hw-stage}), so the weights cannot be streamed from a memory at a warmer temperature stage; they must reside on the chip, next to the logic that uses them. Two implementations satisfy this requirement: programmable SFQ registers, e.g., non-destructive-readout (NDRO) cells or shift-register arrays, which are loaded once per model at a negligible configuration bandwidth, so that one chip can run any model with the same architecture; or weights hardwired into the logic. While hard-wired weights provide a minimum area baseline, a fully reconfigurable deployment requires on-chip NDRO storage. Estimating 12--16 JJs per NDRO bit, storing the 0.18--0.91~Mbit threshold array adds approximately 2.2--14.5~M JJs for memory cells alone. Including full comparators and routing MUXes increases the logic count by 6.8--11$\times$ (Sec.~\ref{sec:res-cost}), bringing the total programmable footprint to $\sim$31--75~M JJs. We adopt hard-wired weights as our baseline to isolate activation logic savings, but note that reconfigurability shifts the primary design cost to on-chip weight storage.

\subsection{Generative Formulation: An Unrolled Non-Equilibrium Transport}
\label{sec:model-flow}

Our main generative model is class-conditional flow matching. We adopt the rectified-flow
interpolation, but instead of the velocity, the network predicts the clean image
$\hat x_1$ and uses a learned per-pixel skip connection:
\begin{equation}
\hat x_1(x,t,c) = \kappa(t)\odot x + f_\theta(x,t,c), \qquad \kappa(t)=\sigma\!\big(W\phi(t)\big),
\end{equation}
where $\phi(t)$ denotes Fourier features of $t$. With $N$ uniform steps, $t_k=k/N$, and
$a_k=\Delta t/(1-t_k)$, one Euler step of the rectified flow becomes
\begin{equation}
\begin{aligned}
x_{k+1} &= \alpha_k\, x_k + \beta_k\, f_\theta(x_k,t_k,c), \\
\alpha_k &= 1-a_k+a_k \kappa(t_k),\quad \beta_k = a_k .
\end{aligned}
\label{eq:euler}
\end{equation}
Because $\alpha_k,\beta_k\ge 0$, Eq.~\eqref{eq:euler} is implemented exactly by one MUX per
pixel (Eq.~\eqref{eq:mux}) with select probability $s=\alpha_k g_x/(\alpha_k g_x+\beta_k
g_f)$ and output gain $g'=\alpha_k g_x+\beta_k g_f$, where $g_x$ and $g_f$ are the gains of
the two inputs. Since $\alpha_k+\beta_k=1+a_k\kappa(t_k)$ and $\kappa\in(0,1)$, the output
gain satisfies
\begin{equation}
g' \le \big(1+a_k\kappa(t_k)\big)\max(g_x,g_f) \le (1+a_k)\max(g_x,g_f).
\label{eq:gainbound}
\end{equation}
The residual stage therefore increases the gain by at most a bounded factor per step. With
$a_k=1/(N-k)$, these factors accumulate over the whole chain to at most
$\prod_{k=0}^{N-1}(1+a_k)=N+1$ relative to the largest incoming gain, i.e., linearly rather
than exponentially in the number of steps. Any further growth stems from the network gain
$g_f$, which depends on $g_x$ through the butterfly layers (Sec.~\ref{sec:model-train}) and
is controlled by the gain regularizer. Predicting the velocity instead would require the
network output itself to represent $(\hat x_1-x_k)/(1-t_k)$, including the cancellation of
$x_k$, which inflates its gain (Sec.~\ref{sec:model-nf}); in our preliminary experiments,
this made the output gain grow to $10^9$ within four steps. The skip connection is merged
into the same MUX and therefore costs no hardware. We use $N=4$ steps and one reflow round, which
straightens the paths sufficiently for four-step sampling.

The $N$ Euler steps are unrolled in hardware: each step has its own replica of the network
and its own random-number generators. A sample is thus produced by a single pass through a
feed-forward pipeline---a finite-time, non-equilibrium transport from noise to data.
Unlike accumulator-based stochastic dynamics, which yield the correct result only after
they have relaxed, every register bit of this pipeline is a fixed function of a finite
set of random draws (Sec.~\ref{sec:model-indep}), so no burn-in is needed.

Both models in this paper generate class-conditional images, but they produce a sample in
different ways. Besides the flow chain, we implement the decoder of a class-conditional
VAE~\cite{kingma2014vae,sohn2015cvae} trained on $8\times8$ digits, which maps a latent
noise vector and the label to an image in a single pass through one network, without
unrolled steps or residual MUXes. A conditional VAE of similar scale, generating MNIST
digits, was also used for the single-chip validation of CN101~\cite{holdijk2026cn101};
we use the decoder as a smaller reference design for the verification, the independence
study, and the cost analysis.

\subsection{Network Architecture: Butterfly Layers}
\label{sec:model-net}

The network $f_\theta$ has four layers of width $n=2^m$. Each layer is a
butterfly-factorized linear map~\cite{dao2019butterfly} followed by a bias, and every layer
except the last is followed by an activation. Throughout the paper, we therefore
distinguish three nested levels: the $N$ Euler \emph{steps}, each with its own replica of
the network; the four \emph{layers} of each replica, three of which carry an activation;
and the $m$ butterfly \emph{stages} of each layer. In the four-step flow chain, a sample
thus passes through 16 layers and 12 activation layers. A butterfly map is the product of $m$ sparse
stages; stage $k$ pairs lane $i$ with lane $j=i\oplus 2^k$ and applies the $2\times2$ block
\begin{equation}
\begin{pmatrix} y_i\\ y_j\end{pmatrix} =
\begin{pmatrix} a & b\\ c & d\end{pmatrix}
\begin{pmatrix} x_i\\ x_j\end{pmatrix}.
\end{equation}
Each layer therefore has $2n\log_2 n$ parameters instead of $n^2$, and every lane feeds
exactly two outputs per stage, which meets the fan-out constraint. The input lanes of the
flow model are arranged as $[x\,|\,\mathrm{onehot}(c)\times4\,|\,0]$, i.e., the pixels,
four copies of the one-hot class label, and zero padding. The copies lie within one network
replica and are unrelated to the four Euler steps. In a butterfly network, each input lane
reaches each output through exactly one path, whose weight is a product of one factor per
stage, so a single copy of the label is diluted by the MUXes, and the gain regularizer
then trades class conditioning for lower gains. Repeating the label removes this
trade-off: for the VAE decoder at the same regularizer strength, the average class
accuracy over three seeds rises from 54--67\% with one copy to 94--96\% with four. The
copies cost no hardware, since
the label lanes are exact 0/1 streams without generators and occupy lanes that would
otherwise be zero padding. Time is injected as a per-step
first-layer bias $W_t\phi(t_k)$, so each unrolled replica has fixed tables.

\subsection{Stochastic Normal Form}
\label{sec:model-nf}

Each butterfly output is a signed combination of two inputs, whereas a MUX computes a
convex combination of two streams. We bridge this gap exactly by propagating the gains
forward through the network. Let $g_i$ be the gain of lane $i$ entering stage $k$, and let
$w_{\mathrm{self}}$ and $w_{\mathrm{partner}}$ be the weights of output $i$ on $x_i$ and
$x_j$, respectively. With $a'=w_{\mathrm{self}}\,g_i$ and $b'=w_{\mathrm{partner}}\,g_j$,
we define
\begin{equation}
G_i = |a'|+|b'|,\qquad s_i=\frac{|a'|}{G_i}.
\label{eq:nf}
\end{equation}
Output $i$ is then implemented by one MUX per channel with select probability $s_i$, whose
inputs are $x_i$ and $x_j$ with their two channels swapped wherever the corresponding
weight is negative. This \emph{sign routing} is pure wiring. The new gain of lane $i$ is
$G_i$. The bias is implemented by an additional MUX that mixes a constant-1 or constant-0
stream into the appropriate channel with select probability $g_i/(g_i+|{\rm bias}_i|)$.
The transformation is exact because it uses precisely the diagonal-scaling redundancy
between butterfly stages: of the $2n\log_2 n$ parameters, $n(\log_2 n-1)$ are redundant,
leaving $n\log_2 n$ select probabilities and $n$ output gains. After training, each $s_i$
is quantized to $B=8$ bits as a $(B+1)$-bit threshold
$S_i=\lfloor s_i 2^B+\tfrac12\rfloor\in\{0,\dots,2^B\}$, so that $s=0$ and $s=1$ are
represented exactly.

The gains carry the scale of the represented values. Since a stream carries $y/g$, a large
gain implies a small probability difference and hence a long readout window for a given
accuracy. Internal cancellation---stages that amplify and then cancel---inflates the
gains. We therefore make the gain propagation of Eq.~\eqref{eq:nf} differentiable and add
a gain regularizer to the loss (Sec.~\ref{sec:model-train}).

\subsection{Activation Function: From a ReLU Counter to a Stateless AND-NOT Gate}
\label{sec:model-act}

A split-unipolar stream represents $y/g=p^+-p^-$, where $p^\pm=\E[z^\pm]$. Because the
gains are positive, $\mathrm{ReLU}(g\,u)=g\,\mathrm{ReLU}(u)$, so a ReLU only needs to
compute $\max(p^+-p^-,0)$ on the stream. 
The standard SC implementation is a per-neuron
saturating up/down counter $D\in[0,2^K-1]$ that acts as a ``debt'' counter: an input
$(1,0)$ pays off one unit of debt if $D>0$ and otherwise emits a 1, whereas an input
$(0,1)$ adds one unit of debt. With an unbounded counter, the long-run output rate is
$P(a\wedge\neg b)-P(\neg a\wedge b)=p^+-p^-$ whenever this quantity is positive, even for
correlated channels. A $K$-bit counter, however, discards the debt that arrives while it is
saturated, so its output rate is $p^+-p^-+P(\neg a\wedge b)\,\pi_{\mathrm{sat}}$, where
$\pi_{\mathrm{sat}}$ is the fraction of cycles at full scale. This bias is non-negative and
largest near $p^+=p^-$, where, for temporally independent inputs, it reaches about
$P(\neg a\wedge b)/2^K$; the counter thus realizes a smoothed ReLU whose transition width
shrinks as $2^{-K}$, consistent with the bias floors measured in
Sec.~\ref{sec:res-explore}. In addition, the counter is stateful: it needs $K$ bits per
neuron, an up/down adder, and a zero test inside a feedback loop, and it has a start-up
transient.
According to our SFQ cost model (Sec.~\ref{sec:hw-cost}), a $K=8$ counter costs
about 3.9\,k JJs and forms an eight-level loop, and the counters together account for
60--70\% of the design.

To eliminate this overhead, we replace the counter with a \emph{stateless} gate on the two
channels:
\begin{equation}
\mathrm{out}(t) = z^+(t) \wedge \neg z^-(t-d_\ell), \qquad
\E[\mathrm{out}] = p^+\,(1-p^-),
\label{eq:andnot}
\end{equation}
with output channels $(\mathrm{out},0)$ and a delay $d_\ell\ge1$ in layer $\ell$. The delay
on $z^-$ is essential. The two channels of a lane are produced by the same MUX selects and
are therefore strongly correlated within a clock cycle, so
$\E[z^+(t)\wedge\neg z^-(t)]$ is not a function of $p^\pm$. In contrast, $z^+(t)$ and
$z^-(t-d_\ell)$ are built from different random draws of the upstream generators, so
Eq.~\eqref{eq:andnot} holds in expectation. Section~\ref{sec:model-indep} gives the precise
conditions: with $d_\ell=1$, Eq.~\eqref{eq:andnot} is exact in the first activation layer,
and with power-of-two delays $d_\ell=1,2,4,\dots$ it is exact in every layer.

Unlike the ReLU, Eq.~\eqref{eq:andnot} depends on $p^+$ and $p^-$ separately, not only on
their difference, and it does not commute with the gains; the network must therefore be
trained on the stream probabilities themselves. When one channel of a lane is close to
zero, which training tends to produce, the gate behaves like $\max(p^+-p^-,0)$. We also
explored $p^+(1-p^-)^2$, obtained with two delayed copies of $z^-$, which brought no
improvement, and a network without any activation, which suffices for the VAE decoder but
suffers from mode collapse on the flow workloads (Sec.~\ref{sec:results}).

\subsection{Independence of the Stochastic Operands}
\label{sec:model-indep}

Eqs.~\eqref{eq:mux} and~\eqref{eq:andnot} hold only if their operands are independent. We
make this requirement precise under an \emph{ideal-RNG model}, in which every stage
generator draws a fresh, uniformly distributed value in every cycle and all generators are
mutually independent. Section~\ref{sec:res-indep} quantifies how closely the implemented
generators follow this model.

\paragraph*{Draw sets} Every pipeline stage is a register with a latency of one cycle, and
all paths through a layer have the same length. Each register bit $z(t)$ is therefore a
deterministic function of a finite set of draws: for every upstream generator $q$, it
depends on the draws $R_q(t-o)$ with offsets $o$ in a set $O_q(z)$ fixed by the pipeline
structure. A chain of MUX stages adds a constant to all offsets, so $|O_q(z)|=1$ for every
generator upstream of the first activation. Constant streams (the label lanes and the bias
inputs) depend on no draw.

\paragraph*{Lemma 1 (MUX)} The select of a MUX in stage $k$ is a function of the current
draw $R_k(t)$ only, whereas its data inputs depend only on draws of generators upstream of
stage $k$, because every stage, and every unrolled replica, owns its generator. The select
is therefore independent of the data, and Eq.~\eqref{eq:mux} is exact. The data inputs may
share draws and be arbitrarily correlated with each other.

\paragraph*{Lemma 2 (AND-NOT)} Let $O_q$ denote the (common) offset set of the two channels
at the input of an activation. The operands $z^+(t)$ and $z^-(t-d)$ depend on the draws
with offsets $O_q$ and $O_q+d$, respectively, which are disjoint for every generator $q$ if
and only if $d\notin O_q-O_q$. In the first activation layer, $|O_q|=1$ and
$O_q-O_q=\{0\}$, so any $d\ge1$ yields independent operands and Eq.~\eqref{eq:andnot} is
exact. An activation with delay $d$, however, outputs a stream that depends on the offsets
$O_q\cup(O_q+d)$. With $d_\ell=1$ in every layer, the second activation therefore sees the
offset sets $\{0,1\}$ and $\{1,2\}$, which share the draws at offset~1, and its operands
are weakly correlated. With power-of-two delays $d_\ell=2^{\ell-1}$, the operands of
activation layer $\ell$ have the offset set $\{0,\dots,2^{\ell-1}-1\}$, whose difference
set does not contain $d_\ell$, and Eq.~\eqref{eq:andnot} is exact in every layer. Both
lemmas require every activation to be stateless; a counter with unbounded memory would make
the draw sets unbounded.

\paragraph*{Design choice} 
The VAE decoder has three activation layers and therefore uses the delays $d_\ell = 1, 2, 4$, at a cost of $d_\ell - 1$ flip-flops per neuron in addition to the inverter (Sec.~\ref{sec:hw-act}). In the flow chain, a sample passes through twelve activation layers in sequence (three in each of the four steps, Sec.~\ref{sec:model-net}), so the power-of-two delays would grow to $2^{11} = 2048$ cycles in the last one; in addition, the residual MUXes merge streams with different offset sets, which enlarges the sets further. We therefore keep $d_\ell = 1$ in the flow chain and accept a small structural covariance in the deeper activations, which we measure in Sec.~\ref{sec:res-indep}. Although this structural covariance remains below $10^{-4}$ in stream probability for $N=4$ steps, its cumulative effect on output distribution bias for deeper chains ($N \ge 8$, $\ge 24$ layers) remains an open theoretical question. Empirical monitoring shows no drift at $N=4$, but scaling to deeper sampler chains may require periodic re-decorrelation stages or reset flip-flops to prevent high-order covariance buildup.

\subsection{Training in Stream Space}
\label{sec:model-train}

\paragraph*{Stream-space forward pass} In addition to the value $Y$ of each lane, we
propagate its gain-scaled \emph{common mode} $C=g\,(p^++p^-)$ and its gain $g$. A MUX
mixes the common modes with the magnitudes $|a'|$ and $|b'|$, and sign routing swaps the
channels but preserves $p^++p^-$. All three quantities therefore follow butterfly
recursions,
\begin{equation}
Y \leftarrow W_k Y,\qquad C \leftarrow |W_k|\,C,\qquad g \leftarrow |W_k|\,g,
\end{equation}
together with the bias terms $Y\mathrel{+}=\mathrm{bias}$, $C\mathrel{+}=|\mathrm{bias}|$,
and $g\mathrel{+}=|\mathrm{bias}|$. The canonical input encoding yields $C=|x|$. The stream
probabilities are recovered as $p^\pm=(C\pm Y)/(2g)$, and applying the activation of
Eq.~\eqref{eq:andnot} to them gives $Y=C=g\,p^+(1-p^-)$. Because $g$ follows the same
recursion as $C$, the gains may differ from sample to sample, as required by the per-step
time bias. This forward pass computes the expectation of the hardware under the
independence model of Sec.~\ref{sec:model-indep}; it matches the separately implemented
golden evaluator to within $10^{-9}$.

\paragraph*{Hardware-consistent common mode} The AND-NOT activation depends not only on
the pixel values but also on the common mode $C$ and the gain $g$ of the incoming pixel
streams. In the hardware, these are produced by the residual MUXes of all preceding steps:
after the first Euler step, the pixel streams are mixtures, and their common mode exceeds
$|x|$. The first training phase, however, samples interpolated states $x_t$ at random
times, which provide the pixel values but not this history. We therefore
\emph{teacher-force} the common mode, in the sense used for sequence models: instead of
the states that the model itself would generate, we use the ground-truth interpolated
states. For a sample in hardware step $k$, we propagate $C$ and $g$ through the residual
MUXes of steps $0,\dots,k-1$, evaluated at the interpolated states $x_{t_j}$, without
gradients. Without this step, the model learned in the first phase is severely mismatched
to the hardware chain; on Fashion-MNIST (Sec.~\ref{sec:results}), for instance, the chain
loss at the start of reflow was $1176$ instead of $1.2$, and the model never recovered.

\paragraph*{Unrolled-chain reflow} In the reflow phase, we train the unrolled $N$-step
chain directly: $x$, $C$, and $g$ are propagated through the residual MUXes exactly as in
the hardware, and the $\hat x_1$ loss is applied at every step. The reflow targets are
generated by a fine-grained ODE sampler whose common mode follows the same hardware rule.

\paragraph*{Gain regularizer} The loss includes $\lambda$ times the mean over all steps of
\begin{equation}
\frac{1}{P}\sum_i \log(1+g_{k,i}) \;+\; \tau\Big(\log\sum_i e^{\log(1+g_{k,i})/\tau} - \log P\Big),
\label{eq:reg}
\end{equation}
where $g_{k,i}$ is the gain of pixel $i$ after step $k$ and $P$ is the number of pixels.
The first term keeps the typical gain small; because $\log(1+g)$ is bounded from below, it
cannot reward switching a neuron off. The second term, a soft maximum with temperature
$\tau$, was added after we observed that the mean term alone allowed a few pixel gains to
diverge (with median gains of 120--174 in some runs), which then dominated both the readout
noise and the quantization error. We use $\tau=0.25$.

\section{Superconducting Accelerator}
\label{sec:hw}

This section describes the accelerator top-down: the system architecture, the pipeline
organization and its verification, the building blocks (butterfly stages, activation cell,
random-number generators, and readout), and finally the SFQ cost and timing model.

\subsection{System Overview}
\label{sec:hw-system}

\begin{figure*}[t]
\centering
\begin{tikzpicture}[
  font=\footnotesize, >={Stealth[length=1.6mm]},
  blk/.style={draw, rounded corners=2pt, minimum height=9mm, align=center, fill=white},
  host/.style={blk, fill=gray!12},
  rng/.style={draw, circle, inner sep=1pt, font=\scriptsize, fill=white},
  every node/.append style={inner sep=3pt}]
  \node[host] (h0) {Host\\encode $x_0$, label};
  \node[blk, right=5mm of h0] (enc) {SNG\\encoder};
  \node[blk, right=6mm of enc, minimum width=24mm] (s0) {Step 0: network\\replica $f_\theta(\cdot,t_0)$};
  \node[blk, right=4mm of s0] (r0) {residual\\MUX};
  \node[right=4mm of r0] (dots) {$\cdots$};
  \node[blk, right=4mm of dots, minimum width=24mm] (s3) {Step $N{-}1$: network\\replica $f_\theta(\cdot,t_{N-1})$};
  \node[blk, right=4mm of s3] (r3) {residual\\MUX};
  \node[blk, right=5mm of r3] (cnt) {readout\\counters};
  \node[host, right=5mm of cnt] (h1) {Host\\$\hat y=g\frac{N^+-N^-}{T}$};
  \draw[->] (h0) -- (enc);
  \draw[->] (enc) -- (s0);
  \draw[->] (s0) -- (r0);
  \draw[->] (r0) -- (dots);
  \draw[->] (dots) -- (s3);
  \draw[->] (s3) -- (r3);
  \draw[->] (r3) -- (cnt);
  \draw[->] (cnt) -- (h1);
  \draw[->] ($(enc.east)+(1.5mm,0)$) |- ++(0,-8mm) -| (r0.south);
  \draw[->] ($(dots.east)+(1mm,0)$) |- ++(0,-8mm) -| (r3.south);
  \node[rng, above=2mm of enc] (ri) {$R_{\mathrm{in}}$};
  \draw[->] (ri) -- (enc);
  \node[rng, above=2mm of r0] (rr0) {$R_{\mathrm{res}}$};
  \draw[->] (rr0) -- (r0);
  \node[rng, above=2mm of r3] (rr3) {$R_{\mathrm{res}}$};
  \draw[->] (rr3) -- (r3);
  \node[blk, below=16mm of s0, minimum width=16mm] (st0) {stage 0\\MUXes};
  \node[right=2mm of st0] (sdots) {$\cdots$};
  \node[blk, right=2mm of sdots, minimum width=16mm] (st1) {stage $m{-}1$\\MUXes};
  \node[blk, right=3mm of st1] (bias) {bias\\MUX};
  \node[blk, right=3mm of bias] (act) {AND-NOT\\$z^+\wedge\neg z^-_{t-1}$};
  \foreach \s in {st0,st1,bias} { \node[rng, above=1.5mm of \s] (q\s) {$R$}; \draw[->] (q\s) -- (\s); }
  \draw[->] (st0) -- (sdots); \draw[->] (sdots) -- (st1); \draw[->] (st1) -- (bias);
  \draw[->] (bias) -- (act);
  \begin{scope}[on background layer]
    \node[draw, dashed, rounded corners, fill=blue!4, fit=(st0)(act)(qst0)(qbias), inner sep=2mm,
          label={[font=\footnotesize]below:one layer ($\times 4$ per replica; no activation after the last)}] (lay) {};
  \end{scope}
  \draw[densely dotted] (s0.south west) -- (lay.north west);
  \draw[densely dotted] (s0.south east) -- (lay.north east);
\end{tikzpicture}
\caption{The proposed accelerator as an unrolled, non-equilibrium flow chain. Each block
is a register stage of bitstream operators with its own random-number generator (circles).
The gains, the input encoding, and the final scaling are handled by the host.}
\label{fig:system}
\end{figure*}
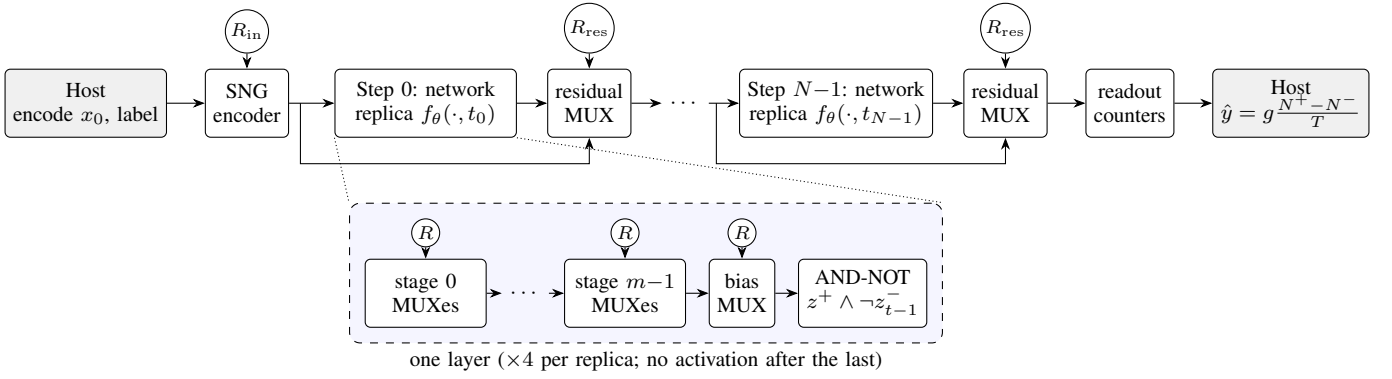

Fig.~\ref{fig:system} shows the unrolled flow chain. The host encodes the initial noise
$x_0$ into $(B+1)$-bit thresholds and drives the label lanes, which are exact constant 0/1
streams and need no generator. On chip, an SNG encoder generates the pixel streams. For
each of the $N$ Euler steps, a replica of the network computes $f_\theta$. Each replica
consists of four butterfly layers, and each layer comprises $m$ MUX stages, a bias stage,
and, except in the last layer, the activation. A residual stage then mixes the network
output with the incoming pixels (Eq.~\eqref{eq:euler}). Readout counters integrate the two
channels of every output pixel over a window of $T$ clock cycles, yielding counts $N^+$ and
$N^-$, from which the host computes $\hat y = g\,(N^+-N^-)/T$. The VAE decoder uses the
same pipeline with a single replica and no residual stage.

Each stage has its own random-number generator (RNG), shared by all MUXes of that stage.
The selects within a stage may be correlated with each other; the only requirement is that
they are independent of the stage's data, which derive from upstream generators.
Replicated datapaths (i.e., the unrolled steps) therefore receive their own RNGs, while
their tables may be shared.

\subsection{Pipeline and Verification Ladder}
\label{sec:hw-pipeline}

The reference implementation is written in synthesizable SystemVerilog with every stage
output registered. Stage $k$ at cycle $t$ combines $R_k(t)$ with data derived from
$R_{k-1}(t-1)$, $R_{k-2}(t-2)$, \dots, so its select never shares a draw with its data
(Lemma~1 in Sec.~\ref{sec:model-indep}); Sec.~\ref{sec:res-indep} verifies this in long
bitstream simulations. A valid shift register gates the readout counters until the
pipeline is full. The four-step digits flow chain comprises 145 register stages. The skip
paths need no delay matching, because the mean of a MUX output depends only on the
independence of its select, not on the alignment of its data inputs.

We verify the design with a ladder of increasingly detailed models: (L0)~the
floating-point PyTorch model~\cite{paszke2019pytorch}; (L1)~the expectation of the normal
form under the independence model, which matches L0 to within $10^{-6}$; (L1q)~the same
model with quantized selects; (L2)~a golden bit-accurate, cycle-accurate simulator, whose
error with respect to L1q decreases as $T^{-1/2}$; and (L3)~the RTL, simulated with
Verilator~\cite{verilator} under cocotb~\cite{cocotb}, which must match L2 \emph{bit for
bit} in every register, activation state, RNG state, and counter in every cycle. All levels
agree for both workloads and both activation functions.

\subsection{Butterfly Stage}
\label{sec:hw-stage}

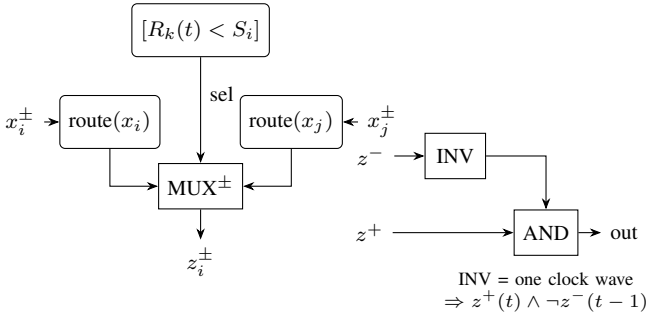
\begin{figure}[t]
\centering
\begin{tikzpicture}[font=\footnotesize, >={Stealth[length=1.5mm]},
  blk/.style={draw, rounded corners=2pt, minimum height=7mm, align=center},
  gate/.style={draw, minimum width=8mm, minimum height=6mm, align=center}]
  \node[blk] (cmp) {$[R_k(t) < S_i]$};
  \node[blk, below=5mm of cmp, xshift=-12mm] (ru) {route$(x_i)$};
  \node[blk, below=5mm of cmp, xshift=12mm] (rw) {route$(x_j)$};
  \node[gate, below=14mm of cmp] (mux) {MUX$^{\pm}$};
  \draw[->] (cmp) -- node[right, pos=0.35] {sel} (mux);
  \draw[->] (ru) |- (mux.west);
  \draw[->] (rw) |- (mux.east);
  \node[left=2mm of ru] (xi) {$x_i^{\pm}$}; \draw[->] (xi) -- (ru);
  \node[right=2mm of rw] (xj) {$x_j^{\pm}$}; \draw[->] (xj) -- (rw);
  \node[below=4mm of mux] (z) {$z_i^{\pm}$}; \draw[->] (mux) -- (z);
  \node[gate, right=24mm of mux, yshift=4mm] (inv) {INV};
  \node[gate, below=4mm of inv, xshift=12mm] (and) {AND};
  \node[left=4mm of inv] (bm) {$z^-$};
  \node[left=16mm of and] (ap) {$z^+$};
  \draw[->] (bm) -- (inv);
  \draw[->] (inv.east) -| (and.north);
  \draw[->] (ap) -- (and.west);
  \node[right=3mm of and] (o) {out}; \draw[->] (and) -- (o);
  \node[below=1mm of and, align=center, font=\scriptsize] {INV = one clock wave\\$\Rightarrow z^+(t)\wedge\neg z^-(t-1)$};
\end{tikzpicture}
\caption{Left: MUX cell of output $i$ in a butterfly stage, consisting of the select
comparator, sign routing, and one MUX per channel. Right: the stateless AND-NOT activation
cell; in an SFQ gate-level pipeline, the inverter delays $z^-$ by exactly one wave.}
\label{fig:cell}
\end{figure}

Each stage consists of $n$ MUX cells and one RNG (Fig.~\ref{fig:cell}). The select of
output $i$ is produced by a comparator $[R_k(t)<S_i]$ between the $B$-bit RNG value and the
threshold of that output. With programmable weights, $S_i$ and the two sign bits are read
from on-chip registers, and the sign routing becomes a pair of small MUXes. In the
hard-wired implementation that we evaluate (Sec.~\ref{sec:model-constraints}), each
threshold $S_i\in\{0,\dots,2^B\}$ is a constant, the comparator reduces to a fixed chain of
AND/OR gates on the RNG bits, and the sign routing is pure wiring. The two channels of a
lane share the same select.

In an SFQ gate-level pipeline, a comparator of depth $\sim B$ feeding a two-level MUX would
normally force the data inputs through an equal number of balancing flip-flops. 
Standard CAD tools enforce full path balancing by inserting D flip-flops along 
the shorter data paths until all incoming arrival times match, which increases JJ counts by up to 40\%. 
Our design does not need this alignment. Only two conditions must hold: (i)~the RNG bits entering a
comparator must belong to the same wave, and (ii)~the select and its complement must reach
both AND branches of a MUX in the same wave, so that the OR gate never merges two pulses.
The data inputs may arrive in any misaligned waves 
without altering the output expectation (Eq.~\eqref{eq:mux}). We exploit this property,
which we call \emph{relaxed path balancing}, by synthesizing each stage as two separate
blocks---a select block ($R\to$ selects) and a MUX block (selects, data $\to$
outputs)---and connecting them without balancing the data against the selects. The data
path of a stage is then only two gate levels deep (AND, OR), while the comparators,
driven by the RNG, operate in parallel. 

\subsection{Activation Cell}
\label{sec:hw-act}

The AND-NOT activation requires $z^-$ to reach the AND gate $d_\ell$ clock waves after
$z^+$. In a gate-level pipelined SFQ circuit, the inverter on $z^-$ is itself a clocked
stage and thus provides exactly one wave of skew, matching $d_\ell=1$ in the golden model.
The cell with $d_\ell=1$ therefore consists of one inverter and one AND gate per neuron,
with no register and no feedback loop (Fig.~\ref{fig:cell}). A delay $d_\ell>1$ adds
$d_\ell-1$ flip-flops on $z^-$; for the decoder delays $1,2,4$
(Sec.~\ref{sec:model-indep}), this amounts to 256 flip-flops, or about 4.6\,k JJs (less
than 1\% of the decoder).

Because the cell relies on this skew, standard synthesis introduces two pitfalls. First,
under full path balancing, the tool pads $z^+$ with a flip-flop to align it with the
inverter output; this removes the skew, and the gate computes $z^+(t)\wedge\neg z^-(t)$, a
different, data-dependent function. Second, an explicit register on $z^-$ in the netlist is
balanced away in the same manner. The relaxed-balancing flow of Sec.~\ref{sec:hw-stage}
preserves the required skew.

\subsection{Random-Number Generators and Readout}
\label{sec:hw-rng}

Each stage generator is a 32-bit xorshift generator~\cite{marsaglia2003xorshift} whose top
$B$ bits form the stage value $R_k(t)$. The seeds are distinct, derived from a master
seed, and loaded at reset; multiple seed sets allow the host to average several
independent windows. In SFQ, the next-state logic consists of three XOR levels (61 XOR
gates and 35 alignment flip-flops, about 2.5\,k JJs). This depth-three feedback loop,
which holds three interleaved generator states, is the only loop remaining in the design.
The readout uses two $W$-bit counters per output pixel, implemented in SFQ as
toggle-flip-flop ripple counters.

\subsection{SFQ Cost and Timing Model}
\label{sec:hw-cost}

To evaluate complete networks without synthesizing hundreds of individual stages, we
developed a gate-level cost model. Each block of an exported network (butterfly, bias, and
residual stages, input SNGs, RNG next-state logic, and activations) is generated as a
structural netlist and verified against the golden bit-level functions. An estimator then
assigns a clock level to each gate and counts the balancing flip-flops, splitters, and
clock-tree cells in two modes: \emph{full} balancing, as performed by a synthesis tool, and
\emph{relaxed} balancing, in which only signals derived from the same RNG are aligned and
the outputs are not padded. We calibrated the estimator against qPALACE~\cite{qpalace}
synthesis results for four butterfly stages, their select/MUX splits, and the activation
cell. The estimator reproduces the MUX blocks exactly and the complete stages to within a
factor of $0.74$--$1.11\times$.

\section{Experimental Results and Discussion}
\label{sec:results}

\subsection{Experimental Setup}
\label{sec:setup}

\paragraph*{Workloads} The conditional VAE decoder is evaluated on the $8\times8$ UCI
digits dataset~\cite{alpaydin1998digits} (1497 training images); it has four layers of
width 64 and a 16-dimensional latent space. The flow chain uses $N=4$ unrolled steps with
four layers per replica and is evaluated on three datasets: the same $8\times8$ digits,
Fashion-MNIST~\cite{xiao2017fashion} average-pooled to $14\times14$ (60\,k training
images), and CIFAR-10~\cite{krizhevsky2009cifar} converted to luminance and average-pooled
to $16\times16$ (50\,k images). The network width is $n=128$ for digits, $n=256$ for
Fashion-MNIST, and $n=512$ for grayscale CIFAR-10.

\paragraph*{Superconductor technology} JJ counts and cell characteristics are based on
the cell library of the MIT Lincoln Laboratory SFQ5ee fabrication
process~\cite{tolpygo2016process}.

\paragraph*{Training} All models are trained in PyTorch~\cite{paszke2019pytorch} with the
Adam optimizer and a cosine learning-rate schedule. The flow models are first trained with
the rectified-flow objective (30 epochs on Fashion-MNIST and CIFAR, 600 on digits) and then
refined by unrolled-chain reflow (15 and 300 epochs, respectively) on pairs generated by a
fine-grained ODE sampler. Gradients are clipped at a norm of 10 for all Fashion-MNIST and
CIFAR runs and for the AND-NOT runs on digits. The selects are quantized to $B=8$ bits.

\paragraph*{Evaluation metrics} The class accuracy of generated samples is measured with a
logistic-regression classifier trained on the real data. Because Inception features are
ill-suited to images this small, we compute the FID~\cite{heusel2017fid} on the hidden
features of a small MLP classifier trained on each dataset; for reference, the FID between
the real training and test images is 0.73 for Fashion-MNIST and 0.55 for CIFAR. The 1-NN
two-sample test reports the leave-one-out accuracy of a nearest-neighbor classifier that
separates real from generated images within each class; a value of 0.5 means the two sets
are indistinguishable. Diversity is the within-class pixel standard deviation of the
generated samples relative to that of the real images. Precision and
recall~\cite{kynkaanniemi2019precision} ($k=5$, in the same classifier features) separate
sample fidelity from coverage of the data. As a sanity check of the feature FID, adding
Gaussian pixel noise or Gaussian blur to real images increases it monotonically on all
datasets (on Fashion-MNIST, e.g., from 0.73 to 4.5 and 34 for noise with $\sigma=0.1$ and
$0.3$, and to 19 and 64 for blur with $\sigma=1$ and 2 pixels); noise mainly lowers
precision, whereas blur mainly lowers recall. All hardware results are reported for the
quantized model expectation (L1q), and the \emph{readout noise} is the predicted RMSE of
the pixel estimates at $T=4096$, computed from the per-pixel gains and stream
probabilities. Unless stated otherwise, results are the mean over three seeds, with the
range in parentheses.

\subsection{Functional Verification}
\label{sec:res-verif}

\begin{figure}[t]
\centering
\includegraphics[width=\columnwidth]{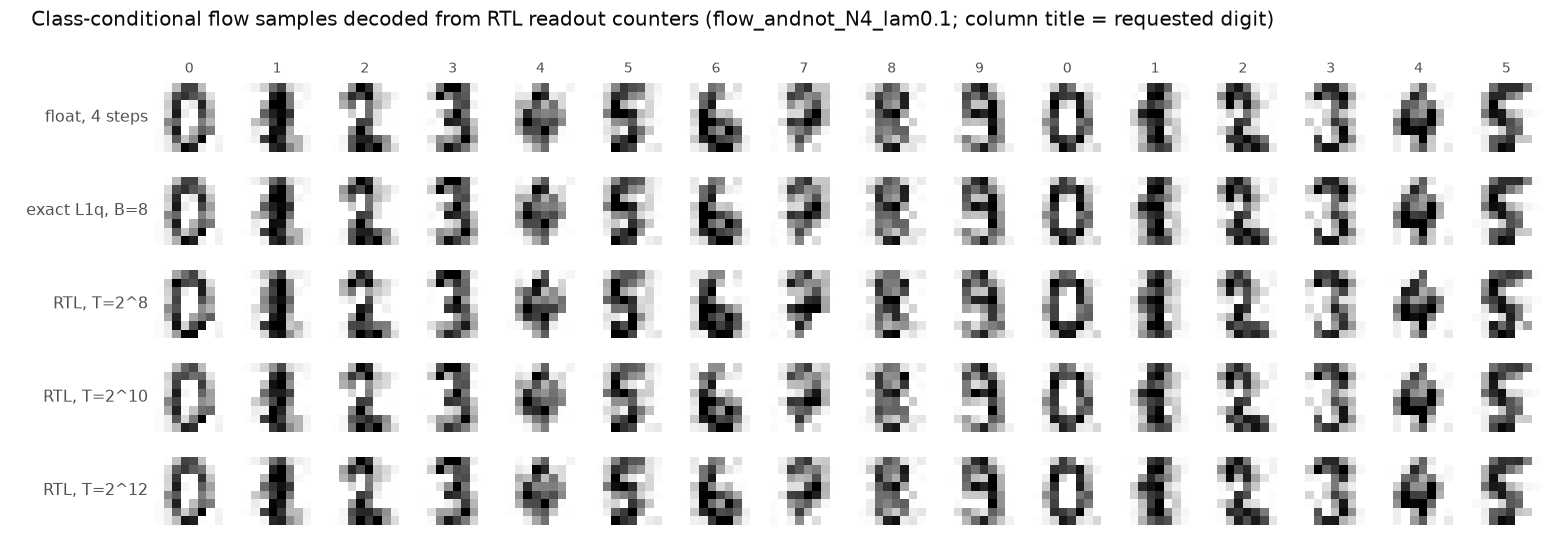}
\includegraphics[width=0.9\columnwidth]{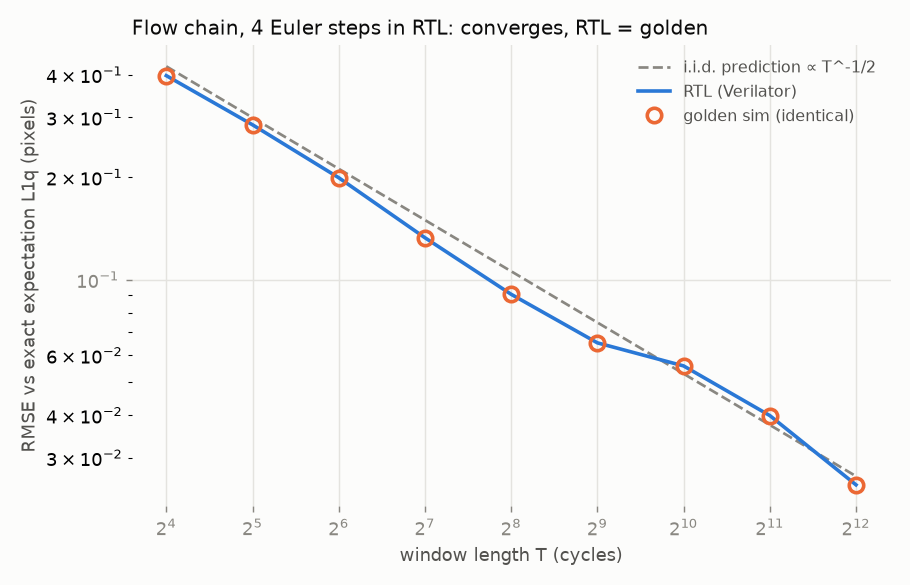}
\caption{Four-step AND-NOT flow chain on $8\times8$ digits ($n=128$). Top: samples decoded
from the RTL readout counters for the requested class (column title). Bottom: error of the
RTL (bit-exact with the golden simulator) with respect to the model expectation (L1q)
versus the window length $T$.}
\label{fig:rtl}
\end{figure}

The RTL implementations of the VAE decoder and of the four-step digits flow chain, each
with both the ReLU and the AND-NOT activation, as well as the decoder with the AND-NOT
delays $1,2,4$, are bit-exact against the golden simulator in every register, activation
state, RNG state, and counter in every cycle, and the readout counters of all test vectors
match at every window length. Fig.~\ref{fig:rtl} shows class-conditional samples decoded
from the RTL counters of the AND-NOT flow chain. Correctly conditioned digits appear from
$T=256$ onward, and the error with respect to the model expectation follows the predicted
$T^{-1/2}$ law, reaching 0.025 at $T=4096$. Unlike the ReLU counter, the AND-NOT activation
has neither a start-up transient nor a bias floor.

\subsection{Independence of the Stochastic Operands}
\label{sec:res-indep}

The lemmas of Sec.~\ref{sec:model-indep} assume ideal generators, whereas the hardware uses
xorshift generators, which are deterministic sequences whose differently seeded instances
are phases of a single recurrence. To quantify how closely the implementation follows the
independence model, we run the golden simulator over long windows ($T=2^{20}$ cycles for
the VAE decoder and $T=2^{18}$ for the digits flow chain, eight test vectors each) and
measure three quantities directly from the bitstreams: for every lane of every MUX, the
covariance difference $\mathrm{Cov}(\mathrm{sel},u)-\mathrm{Cov}(\mathrm{sel},w)$, which
is exactly the bias of Eq.~\eqref{eq:mux}; for every AND-NOT neuron, the covariance
$\mathrm{Cov}(z^+(t),z^-(t-d))$ for $d=0,\dots,8$; and for every register of every layer,
the deviation of its measured rate from the model expectation. Each quantity is converted
into a z-score using an i.i.d.-bit standard error, so that a root-mean-square (rms) z-score
of 1 indicates no detectable deviation. We compare four generator families: the hardware
xorshift generators; ideal i.i.d.\ draws; a distinct full-period xorshift recurrence per
stage; and a distinct maximal-length linear-feedback shift register (LFSR) per stage,
leaped eight steps per cycle. The full periods of the latter two were verified from the
order of their transition matrices over GF(2); the 162 full-period xorshift triples
include the one used in hardware.

\begin{table}[t]
\caption{Independence study on the VAE decoder ($T=2^{20}$, 512 neurons per layer): rms
z-score of the AND-NOT operand covariance at the design's delay, per activation layer,
and of the MUX bias over all stages (1 = not detectable; sampling spread about $\pm0.03$).}
\label{tab:indep}
\centering
\setlength{\tabcolsep}{3pt}
\begin{tabular}{llcccc}
\toprule
Generators & Delays & Act.~1 & Act.~2 & Act.~3 & MUX \\
\midrule
ideal (i.i.d.)        & 1, 1, 1 & 0.93 & 2.91 & 1.90 & 0.95--1.05 \\
xorshift (hardware)   & 1, 1, 1 & 1.02 & 2.83 & 1.87 & 0.92--1.02 \\
distinct xorshift     & 1, 1, 1 & 1.03 & 3.23 & 2.04 & 1.00--1.01 \\
distinct LFSRs        & 1, 1, 1 & 1.01 & 3.10 & 1.80 & 0.97--1.01 \\
\midrule
ideal (i.i.d.)        & 1, 2, 4 & 0.93 & 1.03 & 0.96 & 0.95--1.05 \\
xorshift (hardware)   & 1, 2, 4 & 1.02 & 0.96 & 0.99 & 0.92--1.02 \\
\bottomrule
\end{tabular}
\end{table}

Table~\ref{tab:indep} summarizes the decoder results. First, no MUX shows a detectable
select/data dependence under any generator family (rms z-scores of 0.92--1.05 per layer and
at most 1.18 for any single stage), in agreement with Lemma~1. Second, the first activation
layer shows no detectable correlation, whereas the second and third layers, with the
uniform delay $d_\ell=1$, show the structural covariance predicted by Lemma~2. This
covariance is the same for ideal generators and for all three pseudo-random families, so it
stems from shared upstream draws rather than from generator quality. Its magnitude is about
$10^{-4}$ in stream probability, roughly $40\times$ below the per-stream readout noise at
$T=4096$, and no drift is detectable at the output register even after $2^{20}$ cycles.
Third, the power-of-two delays $1,2,4$ remove the covariance in all layers for both ideal
and xorshift generators, which confirms Lemma~2 and justifies the delays adopted in the
decoder. Without any delay, the same-cycle operands are strongly correlated (rms z-scores
of 76--124), confirming that the delay is essential.

In the digits flow chain, the MUX statistics are again indistinguishable from those
obtained with ideal generators, for which Lemma~1 guarantees exactness (per-stage rms
z-scores of up to 1.43 with ideal and up to 1.51 with xorshift generators; the spread
reflects the small number of independent select draws per stage). With $d_\ell=1$, only the
last activation layer of the last step shows a detectable covariance (rms z-scores of
1.6--1.75 in all seven runs: four xorshift seed sets and three ideal-generator seeds); all
other activation layers are at the noise level at $T=2^{18}$. The register rates of the
input encoder and the residual stages vary between seed sets in the same way for xorshift
and ideal generators, showing that the different seeds of the xorshift recurrence
introduce no measurable defect. In summary, the training forward pass and the L1 evaluator
are exact under the independence model for all MUXes, for the first activation layer, and
for every layer of the decoder; in the deeper layers of the flow chain, they neglect a
measured covariance that lies well below the readout noise.

\subsection{Model Exploration}
\label{sec:res-explore}

\paragraph*{VAE decoder} The decoder with the AND-NOT activation trains successfully (test
negative ELBO of $-15.2$, class accuracy of 0.92) but is worse than the ReLU ($-22.6$,
0.93). A decoder without any activation outperforms both ($-26.5$, 0.94). This workload
thus behaves like a conditional probabilistic PCA and does not need a nonlinearity, so it
cannot be used to rank activation functions. We also found that the ReLU counter needs only
$K=4$ bits: its error at $T=2^{14}$ is essentially the same as with $K=8$ (0.041 vs.\
0.040), whereas $K=2$ leaves a bias floor of about 0.1. The same holds for the ReLU flow
chains on all three datasets: in bit-accurate simulations with the same seeds, the $K=4$
counter increases the error at $T=4096$ by at most 3\% relative to $K=8$, whereas $K=2$
leaves bias floors of 0.03--0.11. We therefore also report the cost of a $K=4$ ReLU in
Sec.~\ref{sec:res-cost}. The flow workloads are the ones in which the nonlinearity matters.

\paragraph*{Flow on digits} Table~\ref{tab:digits} compares the three activation options on
the digits flow chain at matched output gain. Without an activation, the chain suffers from
mode collapse: every class yields nearly identical copies of a single image (diversity of
0.25). Both AND-NOT and ReLU produce varied digits with the same class accuracy; AND-NOT
gives slightly lower diversity but a slightly better 1-NN score.

\begin{table}[t]
\caption{Flow chain on $8\times8$ digits ($n=128$, $N=4$, $\lambda=0.1$, one seed).}
\label{tab:digits}
\centering
\begin{tabular}{lccccc}
\toprule
Activation & Acc. & Div. & 1-NN & Gain & RMSE$_{2^{14}}$ \\
\midrule
none (linear) & 1.000 & 0.25 & 0.962 & 3.2 & 0.013 \\
ReLU counter  & 0.998 & 0.64 & 0.942 & 3.1 & 0.012 \\
AND-NOT       & 1.000 & 0.56 & 0.933 & 3.3 & 0.014 \\
\bottomrule
\end{tabular}
\end{table}

\subsection{Sample Quality versus Hardware Cost}
\label{sec:res-quality}

\begin{figure}[t]
\centering
\includegraphics[width=\columnwidth]{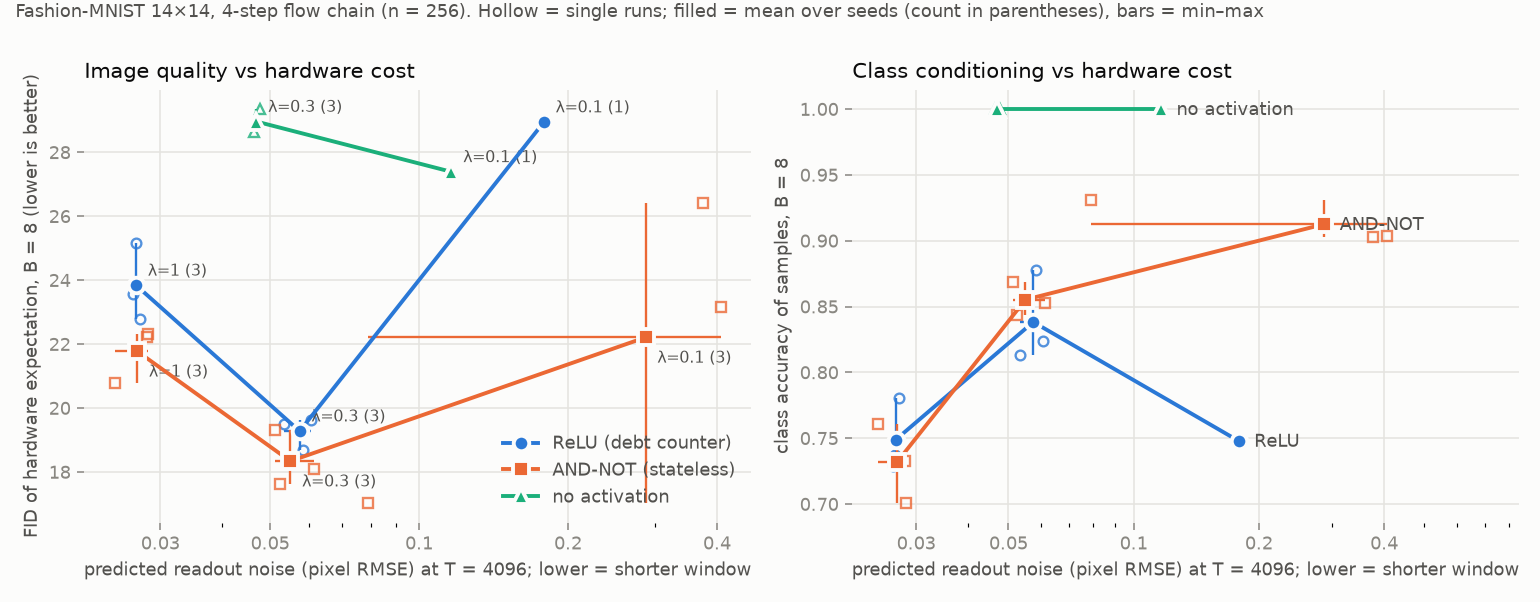}
\caption{Fashion-MNIST: hardware FID and class accuracy versus predicted readout noise for
the mean gain regularizer with $\lambda\in\{0.1,0.3,1\}$ (hollow markers: single runs;
filled markers: seed means with min--max bars). At $\lambda=0.1$, the mean regularizer
alone is unstable; the tail term resolves this (Table~\ref{tab:fashion}).}
\label{fig:pareto}
\end{figure}

\begin{figure*}[t]
\centering
\includegraphics[width=0.325\textwidth]{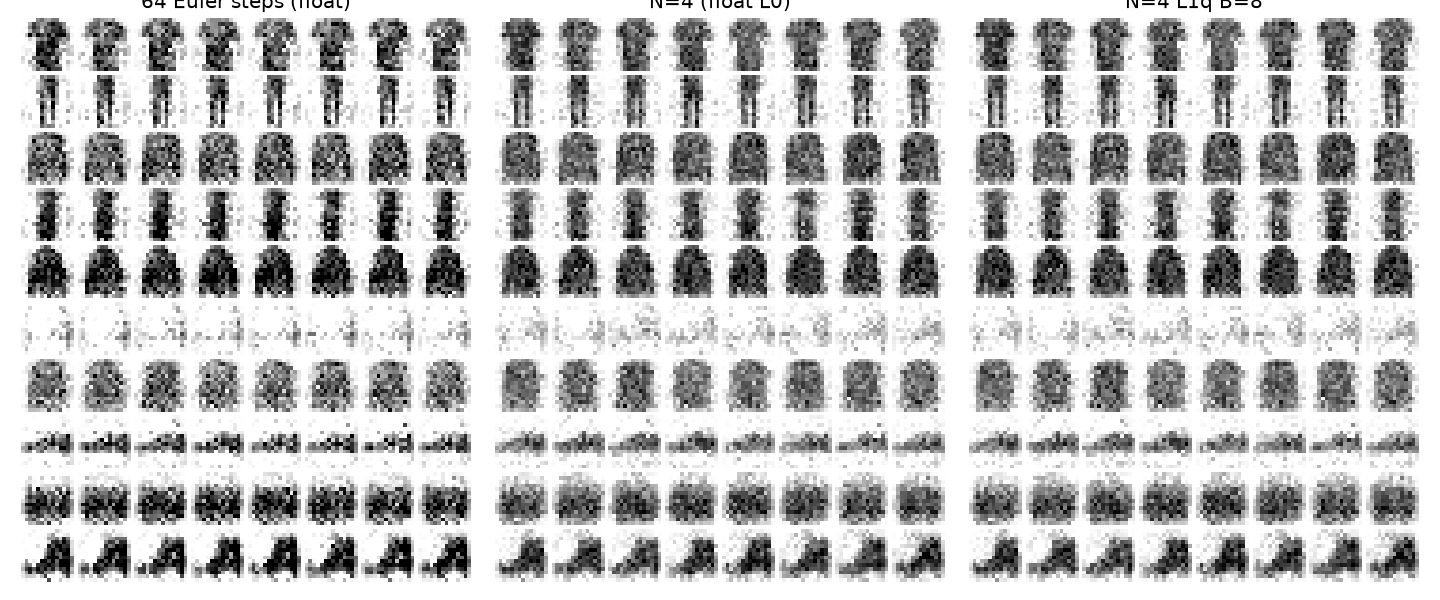}\hfill
\includegraphics[width=0.325\textwidth]{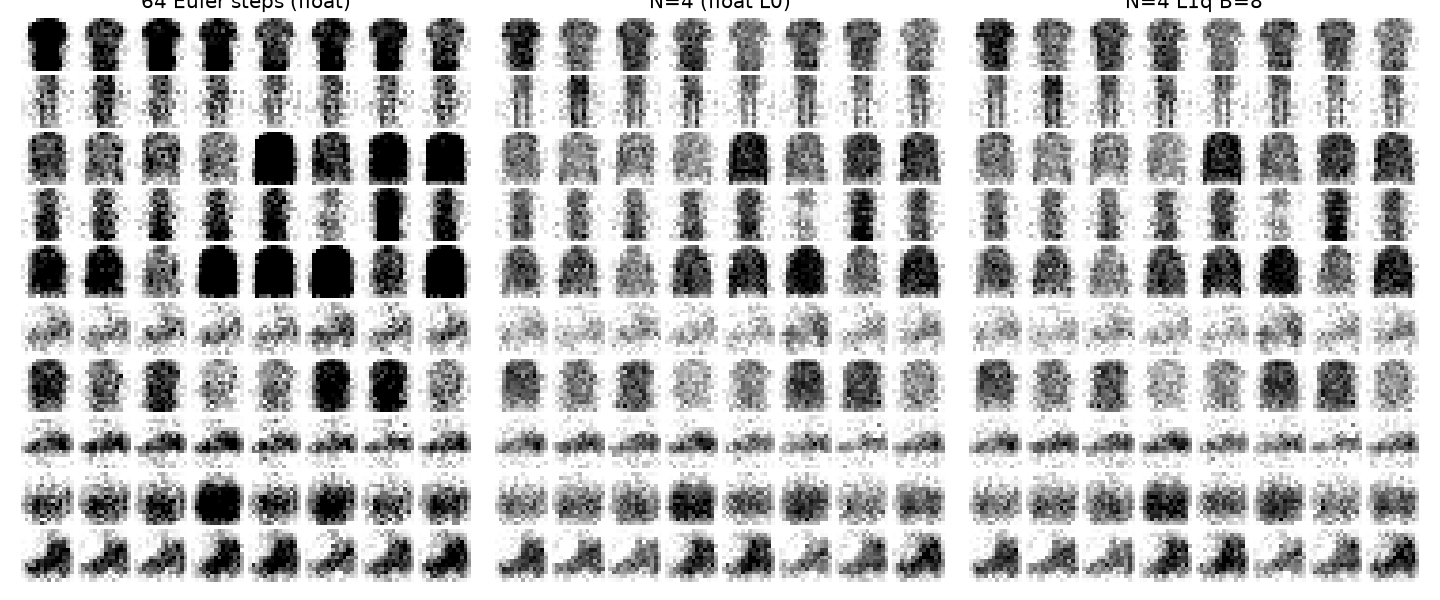}\hfill
\includegraphics[width=0.325\textwidth]{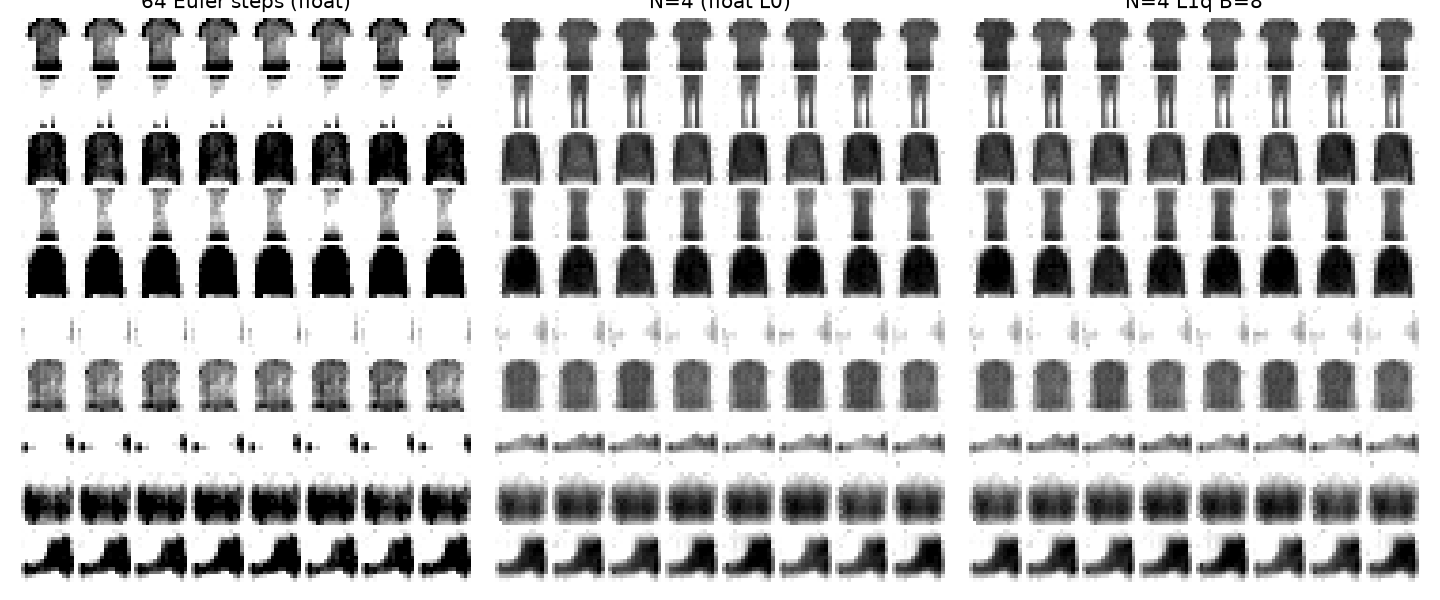}
\caption{Fashion-MNIST samples (one row per class; each panel shows the 64-step
floating-point sampler, the four-step chain, and its quantized hardware expectation).
Left: AND-NOT ($\lambda=0.1$ + tail). Middle: ReLU ($\lambda=0.1$ + tail). Right: no
activation ($\lambda=0.3$), which collapses to a single item per class.}
\label{fig:fashion}
\end{figure*}

The gain regularizer trades sample quality against readout noise, and the readout noise
determines the required window length and hence the time per sample. We therefore compare
the activation functions at matched readout noise. Fig.~\ref{fig:pareto} and
Table~\ref{tab:fashion} report the Fashion-MNIST results. Without an activation, the model
again collapses (FID of 29, diversity of 0.19). At the low-noise end ($\lambda=1$), AND-NOT
clearly outperforms the ReLU (FID of 21.8 vs.\ 23.8, with non-overlapping seed ranges). At
medium noise levels, AND-NOT achieves an FID about one point better and a higher class
accuracy (0.91 vs.\ 0.85 with the tail regularizer), whereas the ReLU samples are more
diverse (0.72 vs.\ 0.60). The tail term of Eq.~\eqref{eq:reg} eliminated the gain runaway
seen at $\lambda=0.1$ with the mean regularizer alone (two of three seeds reached median
gains above 120) and improved the results for both activations. Representative samples are
shown in Fig.~\ref{fig:fashion}.

\begin{table}[t]
\caption{Fashion-MNIST $14\times14$ ($n=256$), three seeds: FID of the quantized hardware
expectation, class accuracy, and readout noise at $T=4096$.}
\label{tab:fashion}
\centering
\setlength{\tabcolsep}{3.5pt}
\begin{tabular}{llccc}
\toprule
Setting & Activation & FID & Acc. & Noise \\
\midrule
\multirow{2}{*}{$\lambda=1$}        & ReLU    & 23.8 (22.8--25.2) & 0.75 & 0.027 \\
                                    & AND-NOT & 21.8 (20.8--22.3) & 0.73 & 0.027 \\
\midrule
\multirow{3}{*}{$\lambda=0.3$}      & none    & 29.0 (28.6--29.4) & 1.00 & 0.047 \\
                                    & ReLU    & 19.3 (18.7--19.6) & 0.84 & 0.057 \\
                                    & AND-NOT & 18.3 (17.6--19.3) & 0.86 & 0.055 \\
\midrule
\multirow{2}{*}{$\lambda=0.1$ + tail} & ReLU  & 18.0 (17.5--18.2) & 0.85 & 0.071 \\
                                    & AND-NOT & 17.4 (16.4--18.0) & 0.91 & 0.079 \\
\bottomrule
\end{tabular}
\end{table}

Grayscale CIFAR-10: CIFAR-10 at $16 \times 16$ grayscale serves as a stress test. A logistic classifier reaches only 27\% accuracy on the real images, and none of our four-layer, four-step models generates recognizable objects. The models nevertheless learn the image statistics (Table~\ref{tab:cifar}). At $n = 512$, the AND-NOT model needs a stronger regularizer: with $\lambda = 0.1$ and the tail term, its median output gain reached 103, and quantization degraded its FID from 10.8 to 15.0. This limitation stems from a dual bottleneck: the sparse $O(n \log n)$ routing of single-instance butterfly layers lacks the channel-mixing capacity required for natural image manifolds, while the aggressive gain regularization ($\lambda=0.3$) necessary to stabilize stream variance restricts expressivity. Generating complex spatial distributions (e.g., full RGB CIFAR-10 or $32\times32$ images) will require cascaded butterfly stages or hybrid residual blocks to increase expressivity without triggering gain runaway.

\begin{table}[t]
\caption{Grayscale CIFAR-10 $16\times16$ ($n=512$) with the tail regularizer, three
seeds (none: one seed).}
\label{tab:cifar}
\centering
\setlength{\tabcolsep}{3.5pt}
\begin{tabular}{lcccc}
\toprule
Configuration & FID & 1-NN & Noise & Gain \\
\midrule
ReLU, $\lambda=0.1$    & 9.42 (9.25--9.66)   & 0.617 & 0.053 & 21.6 \\
ReLU, $\lambda=0.3$    & 11.43 (11.20--11.71) & 0.621 & 0.034 & 9.2 \\
AND-NOT, $\lambda=0.3$ & 11.04 (10.16--12.75) & 0.591 & 0.044 & 5.3 \\
none, $\lambda=0.1$    & 12.61                & 0.686 & 0.096 & 12.4 \\
\bottomrule
\end{tabular}
\end{table}

\subsection{Where Is Sample Quality Lost?}
\label{sec:res-ladder}

To separate the cost of the hardware constraints from the capacity of the model, we
evaluate a ladder of models on each dataset, adding one constraint per rung while keeping
the training recipe, evaluation noise, and metrics fixed
(Table~\ref{tab:ladder})\footnote{All rungs, including the previously reported models,
were re-evaluated with a single evaluation script. Because the MLP classifier that provides
the FID features is retrained for each evaluation and stops at a fixed iteration limit,
its features vary slightly between evaluations, and the resulting FID values can differ by
up to about 0.5 from those in Tables~\ref{tab:fashion} and~\ref{tab:cifar}. Comparisons
within each table are unaffected.}: (A)~a dense MLP of the same width and depth with float
ReLU, using the $\hat x_1$ prediction with the learned skip; (B)~the butterfly network in
float without the gain regularizer; (C)~with the gain regularizer (ReLU); (D)~with the
AND-NOT activation in stream space; (E)~with selects quantized to 8 bits; and (F)~with
bit-level sampling by the golden simulator at $T=4096$. The standard velocity
parameterization is not a suitable float baseline at these widths: the network must carry
the noisy image through its ReLU layers, which needs two lanes per signed pixel, i.e., 392
lanes for Fashion-MNIST at $n=256$. A dense velocity model reaches an FID of only 61 on
Fashion-MNIST and 30 on CIFAR, and 12.4 on Fashion-MNIST at twice the width, which
independently motivates the $\hat x_1$ parameterization of Sec.~\ref{sec:model-flow}.

\begin{table}[t]
\caption{Ablation ladder: FID per rung (mean of three seeds; digits and the bit-level rows:
seed 0). Each rung adds one constraint to the previous one.}
\label{tab:ladder}
\centering
\setlength{\tabcolsep}{3.5pt}
\begin{tabular}{lccc}
\toprule
Rung & Digits & Fashion & CIFAR \\
\midrule
real images (reference)                      & 0.92 & 0.73 & 0.60 \\
A\; dense, float ReLU                         & 1.62 & 3.13 & 2.26 \\
B\; butterfly, float, no regularizer          & 2.17 & 10.3 & 4.40 \\
C\; + gain regularizer (ReLU)                 & 3.07 & 17.0 & 11.2 \\
D\; + AND-NOT, stream space                   & 2.43 & 16.9 & 11.1 \\
E\; + 8-bit selects (L1q)                     & 2.44 & 17.3 & 11.1 \\
F\; + bit level, $T=4096$                     & 2.45 & 17.6 & 11.1 \\
\midrule
ReLU hardware, bit level, $T=4096$            & 3.05 & 19.2 & 11.5 \\
\bottomrule
\end{tabular}
\end{table}

The steps specific to the proposed hardware cost little. Replacing the ReLU by the
stateless AND-NOT activation (C$\to$D) does not increase the FID on any dataset, and
quantizing the selects and sampling at the bit level (D$\to$F) together add less than one
FID point. For the same seed, bit-level sampling at $T=4096$ raises the FID above the
quantized expectation by 0.4 for AND-NOT and 1.5 for the ReLU counter on Fashion-MNIST, and
by 0.8 and 0.0 on CIFAR.

The quality is instead lost in two model-level steps. The butterfly factorization (A$\rightarrow$B) is the area trade-off of the design: it uses 9--28$\times$ fewer weights per layer than the dense network, which would need correspondingly more weight memory and JJs (Sec.~\ref{sec:model-constraints}). The gain regularizer (B$\rightarrow$C) is the price of a finite readout. The unregularized butterfly network reaches its FID only with median stream gains of about $4 \times 10^9$ on Fashion-MNIST, which corresponds to a readout noise of about $10^3$ at $T = 4096$ and would require about $10^{12}$ cycles---minutes---per sample. This loss therefore cannot simply be tuned away: lowering the regularizer of the AND-NOT model to $\lambda = 0.03$ or $0.01$ leaves its floating-point FID essentially unchanged (16.5 and 17.8), while the gains grow by one to three orders of magnitude and both the quantized FID (43 and 72) and the readout noise degrade. Precision and recall analysis reveals that the realizable models lose coverage rather than individual sample fidelity: on Fashion-MNIST, precision drops moderately from 0.99 (Rung A) to 0.82 (Rung D), whereas recall suffers a severe drop from 0.65 to 0.03, initiated primarily by the structural sparsity of the butterfly factorization (dropping to 0.15 at Rung B). Bounding the stream gain further prunes minor generative modes. To mitigate this coverage loss in future work, adaptive temperature annealing on $\tau$ or multi-head butterfly topologies could expand expressivity while maintaining strict gain bounds.

\subsection{SFQ Synthesis and Cost Analysis}
\label{sec:res-cost}

\paragraph*{Relaxed path balancing} Table~\ref{tab:qpalace} reports qPALACE synthesis
results for single butterfly stages of the trained decoder, both fused and split into
select and MUX blocks. Balancing flip-flops account for 41\% of the JJs of a sparse
hard-wired stage. Splitting the stage, so that the data inputs are not aligned with the
selects, reduces the JJ count of a hard-wired stage by 40--47\% and shortens its data path
from eight gate levels to three. Hard-wired weights are also $6$--$8\times$ cheaper than
programmable ones, even before accounting for weight storage.

\begin{table}[t]
\caption{qPALACE synthesis results for one butterfly stage ($n=64$, $B=8$); JJ counts
include the clock tree.}
\label{tab:qpalace}
\centering
\setlength{\tabcolsep}{3pt}
\begin{tabular}{lcccc}
\toprule
Stage & \multicolumn{2}{c}{Balancing DFFs} & \multicolumn{2}{c}{JJs} \\
      & fused & split & fused & split \\
\midrule
hard-wired, sparse (L0, stage 0) & 1055 & 384 & 28.6\,k & 15.3\,k ($-47\%$) \\
hard-wired, dense (L2, stage 5)  & 1069 & 440 & 40.2\,k & 24.2\,k ($-40\%$) \\
programmable                     & 6009 & 4409 & 236\,k & 175\,k ($-26\%$) \\
\bottomrule
\end{tabular}
\end{table}

\paragraph*{Activation cell} qPALACE maps the AND-NOT cell to one inverter and one AND
gate. Under full path balancing, the tool inserts three additional flip-flops per neuron
(102 JJs per neuron, including the clock tree), which also destroys the required skew
(Sec.~\ref{sec:hw-act}); with relaxed balancing, the cell costs about 48 JJs. The ReLU
counter, in comparison, costs 3.9\,k JJs at $K=8$ and 1.7\,k JJs at $K=4$ and contains a
feedback loop.

\paragraph*{Complete networks} Table~\ref{tab:cost} reports the cost-model totals for
hard-wired weights and relaxed balancing. Relative to a ReLU with a $K=8$ counter, the
AND-NOT activation reduces the JJ count by $2.5\times$ for the decoder and by
$3.0$--$3.1\times$ for the flow chains; relative to the sufficient $K=4$ counter
(Sec.~\ref{sec:res-explore}), the reduction is $1.6\times$ for the decoder and $1.9\times$
for both flow chains. Once the activation counters are gone, the butterfly stages dominate
the cost, and the RNGs contain the only remaining feedback loops. With programmable
weights, which require full comparators and routing MUXes, the AND-NOT decoder and digits
chain would require 3.4\,M and 31\,M JJs, respectively ($6.8\times$ and $11\times$ the
hard-wired cost), plus 20\,kbit and 183\,kbit of on-chip weight registers.

\begin{table}[t]
\caption{SFQ cost of complete networks (hard-wired weights, relaxed balancing, qPALACE JJ
accounting) and, for the AND-NOT design, the pipeline latency and time per sample
($T=4096$) at 9\,GHz.}
\label{tab:cost}
\centering
\footnotesize
\setlength{\tabcolsep}{3pt}
\begin{tabular}{lcccc}
\toprule
 & \multicolumn{2}{c}{ReLU} & & \\
Network & $K{=}8$ & $K{=}4$ & AND-NOT & Latency / sample \\
\midrule
VAE decoder ($n{=}64$)     & 1.24\,M & 0.82\,M & 0.50\,M & 7.6\,ns / 0.46\,$\mu$s \\
Flow, digits ($n{=}128$)   & 8.75\,M & 5.36\,M & 2.78\,M & 31.7\,ns / 0.49\,$\mu$s \\
Flow, Fashion ($n{=}256$)  & 18.9\,M & 12.1\,M & 6.2\,M  & 35.2\,ns / 0.49\,$\mu$s \\
\bottomrule
\end{tabular}
\end{table}

\paragraph*{Comparison with a binary design} 
Using the same cost model and clock frequency, we also estimated a fixed-point SFQ implementation of the same networks. It evaluates the same normal form with the same 8-bit tables in 8-bit two’s-complement arithmetic (constant multipliers in canonical signed-digit form, carry-save trees, and Kogge–Stone adders, under full path balancing). A bit-parallel version requires 32–-51$\times$ more JJs than the AND-NOT design (e.g., 321 M instead of 6.2 M JJs for Fashion-MNIST), and a bit-serial version 3.8–-5.9$\times$ more. In return, the binary design produces one sample per clock cycle (bit-serial: one per 17 cycles) instead of one per T = 4096 cycles. This throughput, however, is usable only if every sample enters and leaves the chip at the clock rate. Since memory is scarce in SFQ~\cite{zokaee2021smart}, the samples must be exchanged with memory at a warmer temperature stage, e.g., a cryogenic DRAM at 77 K~\cite{tannu2017cryodram,lee2019cryoram}, which for the Fashion-MNIST chain would require a bandwidth of 28 Tb/s (bit-serial: 1.7 Tb/s). The stochastic design, in contrast, loads its input thresholds and reads its counters only once per window, which requires 21 Gb/s. To transfer these counts off-chip, a parallel-to-serial SFQ readout tree multiplexes the $2P$ output counters ($2 \times 196 = 392$ $W$-bit counters for Fashion-MNIST) into a low-count differential output bus, operating well within the thermal budget of standard 4~K cryo-coolers ($<1$~W dissipation). At equal link bandwidth, the bit-parallel binary design is only about 3$\times$ faster than the stochastic design, while being 51$\times$ larger. In summary, our stateless stochastic design holds stationary thresholds and integrates output counts locally over window $T$, reducing required off-chip bandwidth to just $21\text{ Gb/s}$—a $\sim 1000\times$ reduction that makes sub-terahertz cryogenic AI execution physically realizable within realistic 4~K thermal budgets.

\paragraph*{Timing} 
In a gate-level pipelined SFQ circuit, the clock frequency of the whole accelerator is set by the worst-case stage delay, i.e., the slowest path between two consecutive clocked cells anywhere in the design; the latency of a block only adds to the pipeline fill time. After place and route, interconnect delay dominates, and the critical path lies in the last butterfly stage of each layer, whose MUXes combine lanes that are $n/2$ apart. The remaining logic, including the select comparators, has timing slack. Because the data inputs of a MUX need no alignment (Eq.~\eqref{eq:mux}), a long partner wire can be segmented by a clocked repeater cell without balancing any other path. With one repeater per partner wire, the accelerator operates at 9 GHz, the frequency used throughout this paper. With relaxed balancing, each repeater is a single flip-flop (about 18 JJs, including its share of the clock tree), and the repeaters account for less than 3\% of the JJs of a network, since only the long-wire stage of each layer needs them; with full path balancing, all short paths would instead be padded to match the repeaters. At 9 GHz, the flow chain generates one sample in about 0.49$\mu$s for a window of T = 4096 cycles, of which pipeline fill accounts for only 32–-35 ns (Table~\ref{tab:cost}). Wider networks have longer partner wires and will need several repeaters per wire to sustain this clock frequency. In physical layout, cross-chip $n/2$ routes are driven by Passive Transmission Lines (PTLs) flanked by TX/RX driver pairs ($\sim 30$ JJs per link). Accounting for a PTL phase velocity of $\sim 100\ \mu\text{m/ps}$ in MIT-LL SFQ5ee, a 1--2~mm wire adds $10$--$20$~ps of flight time, which is readily absorbed by two clocked PTL repeater stages at 9~GHz without violating clock setup constraints. Multiphase clocking~\cite{bairamkulov2024multiphase}, proposed to reduce the path-balancing overhead of SFQ circuits, could also be used to spread the delay of long wires over several clock phases, which would relax the critical timing constraint and allow a considerably higher clock frequency. Pulse-level simulation~\cite{delport2019josim} of the laid-out stages is left for future work.

\subsection{Scaling with Network Width}
\label{sec:res-scaling}

\begin{figure*}[t]
\centering
\includegraphics[width=\textwidth]{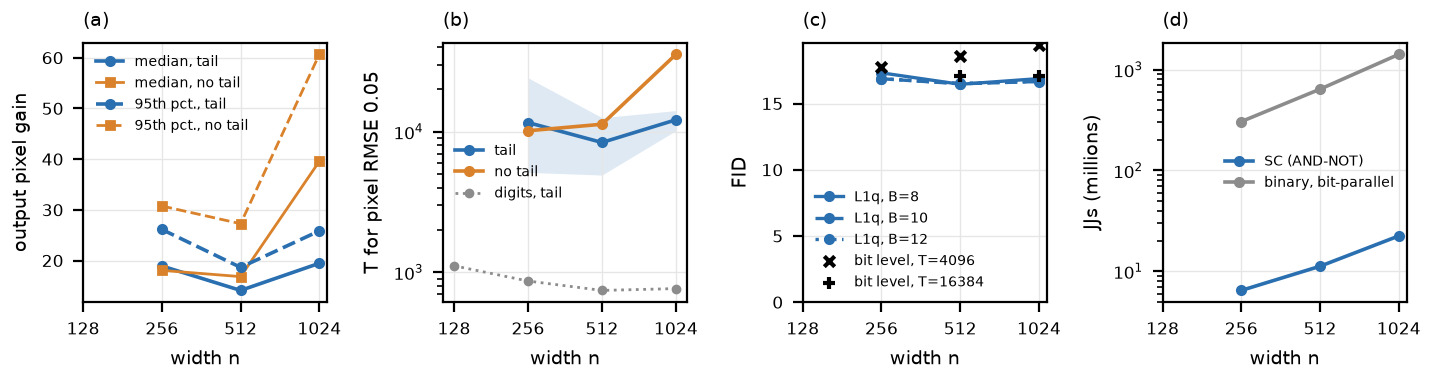}
\caption{Width scaling of the AND-NOT flow chain on Fashion-MNIST ($\lambda=0.1$; three
seeds with the tail term, one seed without it). (a)~Median and 95th-percentile output pixel
gain. (b)~Window $T_{\mathrm{req}}$ required for a pixel RMSE of 0.05 (band: seed range);
digits shown for reference. (c)~FID of the quantized expectation with $B=8$, 10, and 12
select bits, and of bit-level sampling at $T=4096$ and $T=16384$ (seed 0; $n=1024$: 200
samples). (d)~JJ count of the stochastic design and of the bit-parallel binary design
(Sec.~\ref{sec:res-cost}).}
\label{fig:scaling}
\end{figure*}

To determine how the gains, the readout window, the quality, and the cost evolve as the
network grows, and where the approach breaks down, we trained the AND-NOT flow chain on
Fashion-MNIST at widths $n\in\{256,512,1024\}$ (smaller widths cannot hold the 196 pixels
and 40 label lanes) with $\lambda\in\{0.1,0.3,1\}$ and the tail term, and at $\lambda=0.1$
without it. For each model, we compute from the per-pixel gains and stream probabilities
the window $T_{\mathrm{req}}$ that yields a pixel RMSE of 0.05, and the corresponding time
per sample at 9\,GHz (Fig.~\ref{fig:scaling}). With $\lambda=0.1$ and the tail term (three
seeds), the median output gain stays between 14 and 20, $T_{\mathrm{req}}$ stays at
$0.8$--$1.2\times10^4$ cycles (0.95--1.4\,$\mu$s per sample), and the FID stays at
16.5--17.3, while the JJ count grows as $n\log n$, from 6.5\,M to 22\,M. Within this range,
the readout window therefore does not erode the advantage of the design: compared with the
binary design of Sec.~\ref{sec:res-cost} at the same link bandwidth, whose JJ count grows
in the same manner, the stochastic design achieves a $5$--$9\times$ smaller product of JJ
count and time per sample at every width, with no crossover.

The tail term is what keeps the gains bounded: without it, the median gain at $n=1024$
doubles (to 40) and $T_{\mathrm{req}}$ triples ($3.6\times10^4$ cycles, 4.0\,$\mu$s),
whereas at $n\le512$ the difference is small. The limit of the approach lies in the noise
budget rather than in the gains: shortening the window with a stronger regularizer degrades
the quality increasingly with width ($\lambda=0.3$ yields an FID of 18 at $n=256$ and 512
but 32 at $n=1024$, and $\lambda=1$ collapses at $n=512$). Quantization is not a
bottleneck: 10 or 12 select bits improve the FID by at most 0.4. At $T=4096$, the
bit-level FID exceeds that of the quantized expectation by 0.4, 2.1, and 2.6 points at
$n=256$, 512, and 1024, respectively. This gap is readout noise rather than bias, since it
shrinks with longer windows (to 0.6 and 0.2 points at $T=16384$ for $n=512$ and 1024),
which confirms $T_{\mathrm{req}}$ as the appropriate time budget. On digits, the gains and
$T_{\mathrm{req}}$ likewise remain constant up to $n=1024$, but the FID degrades at
$n=1024$ (7.6 instead of 2.3--2.6), where the 64 pixels occupy only a small fraction of the
lanes.

\subsection{Discussion}
\label{sec:discussion}

\paragraph*{Benefits of the stateless activation} Across the three datasets and at matched
readout noise, the AND-NOT activation achieves sample quality comparable to that of the
ReLU: it performs on par on digits, achieves an equal or better FID on Fashion-MNIST, and
is about 0.6 FID points worse than the interpolated ReLU on grayscale CIFAR, while giving a
better 1-NN score on digits and CIFAR. At the same time, it requires $1.6$--$1.9\times$
fewer JJs than a ReLU with a 4-bit counter ($2.5$--$3.1\times$ fewer than with an 8-bit
counter) and eliminates all per-neuron state and feedback loops. These benefits come at
the cost of lower sample diversity than the ReLU and a stronger dependence on gain control
as the network grows.

\paragraph*{Gain control as the scaling constraint} The stream gains shape all results in
this paper, since they determine the readout noise and the degradation due to
quantization. On Fashion-MNIST, the tail regularizer keeps the gains bounded up to
$n=1024$ (Sec.~\ref{sec:res-scaling}), whereas on grayscale CIFAR at $n=512$ the AND-NOT
model required a stronger regularizer. Because the residual stages bound their own
contribution to the gain (Eq.~\eqref{eq:gainbound}), the growth observed without
regularization originates in the network layers. The ablation ladder (Sec.~\ref{sec:res-ladder})
shows that bounding the gains, rather than the stochastic implementation, accounts for most
of the gap to an unconstrained float model. Gain-aware, better-conditioned
parameterizations and a larger $B$ are therefore natural directions for future work.

\section{Conclusion}
\label{sec:conclusion}

We presented a model--hardware co-design of generative AI for superconducting SFQ logic.
The approach follows the spirit of thermodynamic computing---values are carried by
stochastic dynamics and read out as time averages---but, unlike relaxation-based
designs, its datapath holds no state and needs no burn-in: a four-step flow-matching
sampler is unrolled into a
pipeline of multiplexers operating on stochastic bitstreams, and the butterfly layers are
rewritten exactly into an adder-free normal form with stationary weights and a small
weight memory. By replacing the ReLU
counter with a stateless AND-NOT gate, whose one-clock skew SFQ technology provides
inherently, and by training the network on the expected stream statistics of the hardware,
the design achieves ReLU-level sample quality with $1.6$--$3.1\times$ fewer Josephson
junctions and no feedback loops other than those of the random-number generators.
Compared with a bit-parallel binary fixed-point implementation of the same networks, it is
$32$--$51\times$ smaller and needs about three orders of magnitude less off-chip bandwidth,
at the cost of a longer time per sample. Stochastic streams also tolerate relaxed path
balancing, which reduces the JJ count of a stage by up to 47\%. We proved that the MUX
operations are exact under ideal random-number generators and that power-of-two delays
make the stateless activation exact in every layer, and we measured the residual
covariances of the implemented generators to be well below the readout noise. The design
is verified bit-exactly from the golden model down to the RTL. Future work includes
pulse-level simulation and timing closure, the integration of physical randomness, and
improved gain control for larger models.

\bibliographystyle{IEEEtran}
\bibliography{refs}

@misc{holdijk2026cn101,
  author       = {Holdijk, Lars and Melanson, Denis and Mensch, Zier and Birchall, Brandon and
                  Cheung, Vincent and Lehrter, Nicholas and Aifer, Maxwell and Duffield, Samuel and
                  Ernst, Jan Ole and Salegame, Rajath and Martinez, Antonio J. and Crooks, Gavin and
                  Cheng, Miranda and Belateche, Zach and Bright, Marc and Coles, Patrick J. and
                  Sbahi, Faris},
  title        = {{CN101} -- A Digital Thermodynamic Computer for Generative {AI}},
  year         = {2026},
  eprint       = {2608.00754},
  archivePrefix= {arXiv},
  primaryClass = {cs.ET}
}

@misc{conte2019thermodynamic,
  author       = {Conte, Tom and DeBenedictis, Erik and Ganesh, Natesh and Hylton, Todd and
                  Strachan, John Paul and Williams, R. Stanley and others},
  title        = {Thermodynamic Computing},
  year         = {2019},
  eprint       = {1911.01968},
  archivePrefix= {arXiv},
  note         = {Report of the CCC workshop on thermodynamic computing}
}

@misc{coles2023thermodynamic,
  author       = {Coles, Patrick J. and Szczepanski, Collin and Melanson, Denis and
                  Donatella, Kaelan and Martinez, Antonio J. and Sbahi, Faris},
  title        = {Thermodynamic {AI} and the Fluctuation Frontier},
  year         = {2023},
  eprint       = {2302.06584},
  archivePrefix= {arXiv}
}

@inproceedings{sohldickstein2015nonequilibrium,
  author    = {Sohl-Dickstein, Jascha and Weiss, Eric and Maheswaranathan, Niru and Ganguli, Surya},
  title     = {Deep Unsupervised Learning using Nonequilibrium Thermodynamics},
  booktitle = {Proceedings of the 32nd International Conference on Machine Learning (ICML)},
  series    = {PMLR},
  volume    = {37},
  pages     = {2256--2265},
  year      = {2015}
}

@incollection{gaines1969stochastic,
  author    = {Gaines, Brian R.},
  title     = {Stochastic Computing Systems},
  booktitle = {Advances in Information Systems Science},
  editor    = {Tou, Julius T.},
  volume    = {2},
  pages     = {37--172},
  publisher = {Plenum Press},
  year      = {1969}
}

@article{alaghi2013survey,
  author  = {Alaghi, Armin and Hayes, John P.},
  title   = {Survey of Stochastic Computing},
  journal = {ACM Transactions on Embedded Computing Systems},
  volume  = {12},
  number  = {2s},
  pages   = {92:1--92:19},
  year    = {2013}
}

@article{brown2001stochastic,
  author  = {Brown, Bradley D. and Card, Howard C.},
  title   = {Stochastic Neural Computation {I}: Computational Elements},
  journal = {IEEE Transactions on Computers},
  volume  = {50},
  number  = {9},
  pages   = {891--905},
  year    = {2001}
}

@article{marsaglia2003xorshift,
  author  = {Marsaglia, George},
  title   = {Xorshift {RNG}s},
  journal = {Journal of Statistical Software},
  volume  = {8},
  number  = {14},
  pages   = {1--6},
  year    = {2003}
}

@inproceedings{dao2019butterfly,
  author    = {Dao, Tri and Gu, Albert and Eichhorn, Matthew and Rudra, Atri and R{\'e}, Christopher},
  title     = {Learning Fast Algorithms for Linear Transforms Using Butterfly Factorizations},
  booktitle = {Proceedings of the 36th International Conference on Machine Learning (ICML)},
  series    = {PMLR},
  volume    = {97},
  pages     = {1517--1527},
  year      = {2019}
}

@inproceedings{kingma2014vae,
  author    = {Kingma, Diederik P. and Welling, Max},
  title     = {Auto-Encoding Variational {B}ayes},
  booktitle = {International Conference on Learning Representations (ICLR)},
  year      = {2014}
}

@inproceedings{sohn2015cvae,
  author    = {Sohn, Kihyuk and Lee, Honglak and Yan, Xinchen},
  title     = {Learning Structured Output Representation using Deep Conditional Generative Models},
  booktitle = {Advances in Neural Information Processing Systems (NeurIPS)},
  volume    = {28},
  year      = {2015}
}

@inproceedings{ho2020ddpm,
  author    = {Ho, Jonathan and Jain, Ajay and Abbeel, Pieter},
  title     = {Denoising Diffusion Probabilistic Models},
  booktitle = {Advances in Neural Information Processing Systems (NeurIPS)},
  volume    = {33},
  pages     = {6840--6851},
  year      = {2020}
}

@inproceedings{song2021sde,
  author    = {Song, Yang and Sohl-Dickstein, Jascha and Kingma, Diederik P. and Kumar, Abhishek and
               Ermon, Stefano and Poole, Ben},
  title     = {Score-Based Generative Modeling through Stochastic Differential Equations},
  booktitle = {International Conference on Learning Representations (ICLR)},
  year      = {2021}
}

@inproceedings{lipman2023flow,
  author    = {Lipman, Yaron and Chen, Ricky T. Q. and Ben-Hamu, Heli and Nickel, Maximilian and Le, Matt},
  title     = {Flow Matching for Generative Modeling},
  booktitle = {International Conference on Learning Representations (ICLR)},
  year      = {2023}
}

@inproceedings{liu2023rectified,
  author    = {Liu, Xingchao and Gong, Chengyue and Liu, Qiang},
  title     = {Flow Straight and Fast: Learning to Generate and Transfer Data with Rectified Flow},
  booktitle = {International Conference on Learning Representations (ICLR)},
  year      = {2023}
}

@inproceedings{heusel2017fid,
  author    = {Heusel, Martin and Ramsauer, Hubert and Unterthiner, Thomas and Nessler, Bernhard and
               Hochreiter, Sepp},
  title     = {{GANs} Trained by a Two Time-Scale Update Rule Converge to a Local {N}ash Equilibrium},
  booktitle = {Advances in Neural Information Processing Systems (NeurIPS)},
  volume    = {30},
  year      = {2017}
}

@inproceedings{kynkaanniemi2019precision,
  author    = {Kynk{\"a}{\"a}nniemi, Tuomas and Karras, Tero and Laine, Samuli and Lehtinen, Jaakko and
               Aila, Timo},
  title     = {Improved Precision and Recall Metric for Assessing Generative Models},
  booktitle = {Advances in Neural Information Processing Systems (NeurIPS)},
  volume    = {32},
  year      = {2019}
}

@misc{alpaydin1998digits,
  author       = {Alpaydin, Ethem and Kaynak, Cenk},
  title        = {Optical Recognition of Handwritten Digits},
  howpublished = {UCI Machine Learning Repository},
  year         = {1998},
  doi          = {10.24432/C50P49}
}

@misc{xiao2017fashion,
  author       = {Xiao, Han and Rasul, Kashif and Vollgraf, Roland},
  title        = {{Fashion-MNIST}: a Novel Image Dataset for Benchmarking Machine Learning Algorithms},
  year         = {2017},
  eprint       = {1708.07747},
  archivePrefix= {arXiv}
}

@techreport{krizhevsky2009cifar,
  author      = {Krizhevsky, Alex},
  title       = {Learning Multiple Layers of Features from Tiny Images},
  institution = {University of Toronto},
  year        = {2009}
}

@article{likharev1991rsfq,
  author  = {Likharev, Konstantin K. and Semenov, Vasilii K.},
  title   = {{RSFQ} Logic/Memory Family: A New {J}osephson-Junction Technology for Sub-Terahertz-Clock-Frequency Digital Systems},
  journal = {IEEE Transactions on Applied Superconductivity},
  volume  = {1},
  number  = {1},
  pages   = {3--28},
  year    = {1991}
}

@article{tolpygo2016process,
  author  = {Tolpygo, Sergey K. and Bolkhovsky, Vladimir and Weir, Terence J. and Wynn, Alex and
             Oates, Daniel E. and Johnson, Leonard M. and Gouker, Mark A.},
  title   = {Advanced Fabrication Processes for Superconducting Very Large-Scale Integrated Circuits},
  journal = {IEEE Transactions on Applied Superconductivity},
  volume  = {26},
  number  = {3},
  pages   = {1--10},
  year    = {2016}
}

@article{pasandi2019pbmap,
  author  = {Pasandi, Ghasem and Pedram, Massoud},
  title   = {{PBMap}: A Path Balancing Technology Mapping Algorithm for Single Flux Quantum Logic Circuits},
  journal = {IEEE Transactions on Applied Superconductivity},
  volume  = {29},
  number  = {4},
  pages   = {1--14},
  year    = {2019}
}

@article{qpalace,
  title     = {Results from the {ColdFlux} Superconductor Integrated Circuit Design Tool Project},
  author    = {Fourie, Coenrad J. and Jackman, Kyle and Delport, Johannes and Schindler, Lieze and
               Hall, Tessa and Febvre, Pascal and Iwanikow, Lucas and Chen, Olivia and
               Ayala, Christopher L. and Yoshikawa, Nobuyuki and others},
  journal   = {IEEE Transactions on Applied Superconductivity},
  volume    = {33},
  number    = {8},
  pages     = {1--26},
  year      = {2023},
  publisher = {IEEE}
}

@article{delport2019josim,
  author  = {Delport, Johannes A. and Jackman, Kyle and le Roux, Paul and Fourie, Coenrad J.},
  title   = {{JoSIM}---Superconductor {SPICE} Simulator},
  journal = {IEEE Transactions on Applied Superconductivity},
  volume  = {29},
  number  = {5},
  pages   = {1--5},
  year    = {2019}
}

@inproceedings{ishida2020supernpu,
  author    = {Ishida, Koki and Byun, Ilkwon and Nagaoka, Ikki and Fukumitsu, Kosuke and Tanaka, Masamitsu and
               Kawakami, Satoshi and Tanimoto, Teruo and Ono, Takatsugu and Kim, Jangwoo and Inoue, Koji},
  title     = {{SuperNPU}: An Extremely Fast Neural Processing Unit Using Superconducting Logic Devices},
  booktitle = {Proceedings of the 53rd IEEE/ACM International Symposium on Microarchitecture (MICRO)},
  pages     = {58--72},
  year      = {2020}
}

@inproceedings{bairamkulov2024multiphase,
  author    = {Bairamkulov, Rassul and De Micheli, Giovanni},
  title     = {Towards Multiphase Clocking in Single-Flux Quantum Systems},
  booktitle = {Proceedings of the 29th Asia and South Pacific Design Automation Conference (ASP-DAC)},
  year      = {2024},
  note      = {Also available as arXiv:2403.05884}
}

@misc{verilator,
  author       = {Snyder, Wilson},
  title        = {Verilator},
  howpublished = {\url{https://www.veripool.org/verilator}},
  note         = {Version 5.020}
}

@misc{cocotb,
  title        = {cocotb: a Coroutine-Based Cosimulation Testbench Environment for Verifying {VHDL} and {SystemVerilog} {RTL} using {Python}},
  howpublished = {\url{https://www.cocotb.org}},
  note         = {Version 1.9}
}

@inproceedings{paszke2019pytorch,
  author    = {Paszke, Adam and others},
  title     = {{PyTorch}: An Imperative Style, High-Performance Deep Learning Library},
  booktitle = {Advances in Neural Information Processing Systems (NeurIPS)},
  volume    = {32},
  year      = {2019}
}

@inproceedings{tannu2017cryodram,
  author    = {Tannu, Swamit S. and Carmean, Douglas M. and Qureshi, Moinuddin K.},
  title     = {Cryogenic-{DRAM} Based Memory System for Scalable Quantum Computers: A Feasibility Study},
  booktitle = {Proceedings of the International Symposium on Memory Systems (MEMSYS)},
  pages     = {189--195},
  year      = {2017},
  doi       = {10.1145/3132402.3132436}
}

@inproceedings{lee2019cryoram,
  author    = {Lee, Gyu-Hyeon and Min, Dongmoon and Byun, Ilkwon and Kim, Jangwoo},
  title     = {Cryogenic Computer Architecture Modeling with Memory-Side Case Studies},
  booktitle = {Proceedings of the 46th International Symposium on Computer Architecture (ISCA)},
  pages     = {774--787},
  year      = {2019}
}

@inproceedings{zokaee2021smart,
  author    = {Zokaee, Farzaneh and Jiang, Lei},
  title     = {{SMART}: A Heterogeneous Scratchpad Memory Architecture for Superconductor {SFQ}-based Systolic {CNN} Accelerators},
  booktitle = {Proceedings of the 54th Annual IEEE/ACM International Symposium on Microarchitecture (MICRO)},
  pages     = {912--924},
  year      = {2021},
  doi       = {10.1145/3466752.3480041}
}

@article{fu2022jbnn,
  author  = {Fu, Rongliang and Huang, Junying and Wu, Haibin and Ye, Xiaochun and Fan, Dongrui and Ho, Tsung-Yi},
  title   = {{JBNN}: A Hardware Design for Binarized Neural Networks Using Single-Flux-Quantum Circuits},
  journal = {IEEE Transactions on Computers},
  volume  = {71},
  number  = {12},
  pages   = {3203--3214},
  year    = {2022},
  doi     = {10.1109/TC.2022.3215085}
}

@inproceedings{gonzalezguerrero2022usfq,
  author    = {Gonzalez-Guerrero, Patricia and Bautista, Meriam Gay and Lyles, Darren and Michelogiannakis, George},
  title     = {Temporal and {SFQ} Pulse-Streams Encoding for Area-Efficient Superconducting Accelerators},
  booktitle = {Proceedings of the 27th ACM International Conference on Architectural Support for Programming Languages and Operating Systems (ASPLOS)},
  pages     = {963--976},
  year      = {2022},
  doi       = {10.1145/3503222.3507765}
}

@inproceedings{cai2019aqfpsc,
  author    = {Cai, Ruizhe and Ren, Ao and Chen, Olivia and Liu, Ning and Ding, Caiwen and Qian, Xuehai and
               Han, Jie and Luo, Wenhui and Yoshikawa, Nobuyuki and Wang, Yanzhi},
  title     = {A Stochastic-Computing Based Deep Learning Framework Using Adiabatic Quantum-Flux-Parametron Superconducting Technology},
  booktitle = {Proceedings of the 46th International Symposium on Computer Architecture (ISCA)},
  pages     = {567--578},
  year      = {2019},
  doi       = {10.1145/3307650.3322270}
}

@inproceedings{li2023superbnn,
  author    = {Li, Zhengang and Yuan, Geng and Yamauchi, Tomoharu and Zabihi, Masoud and Xie, Yanyue and
               Dong, Peiyan and Tang, Xulong and Yoshikawa, Nobuyuki and Tiwari, Devesh and Wang, Yanzhi and
               Chen, Olivia},
  title     = {{SupeRBNN}: Randomized Binary Neural Network Using Adiabatic Superconductor {J}osephson Devices},
  booktitle = {Proceedings of the 56th Annual IEEE/ACM International Symposium on Microarchitecture (MICRO)},
  pages     = {584--598},
  year      = {2023},
  doi       = {10.1145/3613424.3623771}
}

@article{crotty2010jjneurons,
  author  = {Crotty, Patrick and Schult, Daniel and Segall, Ken},
  title   = {{J}osephson Junction Simulation of Neurons},
  journal = {Physical Review E},
  volume  = {82},
  pages   = {011914},
  year    = {2010}
}

@article{schneider2020synaptic,
  author  = {Schneider, Michael L. and Donnelly, Christine A. and Haygood, Ian W. and Wynn, Alex and
             Russek, Stephen E. and Castellanos-Beltran, Manuel and Dresselhaus, Paul D. and
             Hopkins, Peter F. and Pufall, Matthew R. and Rippard, William H.},
  title   = {Synaptic Weighting in Single Flux Quantum Neuromorphic Computing},
  journal = {Scientific Reports},
  volume  = {10},
  pages   = {934},
  year    = {2020}
}

@misc{bozbey2020sfqsnn,
  author       = {Bozbey, Ali and Karamuftuoglu, Mustafa Altay and Razmkhah, Sasan and Ozbayoglu, Murat},
  title        = {Single Flux Quantum Based Ultrahigh Speed Spiking Neuromorphic Processor Architecture},
  year         = {2018},
  eprint       = {1812.10354},
  archivePrefix= {arXiv}
}

@article{karamuftuoglu2023jjsoma,
  author  = {Karamuftuoglu, Mustafa Altay and Bozbey, Ali and Razmkhah, Sasan},
  title   = {{JJ-Soma}: Toward a Spiking Neuromorphic Processor Architecture},
  journal = {IEEE Transactions on Applied Superconductivity},
  volume  = {33},
  number  = {8},
  pages   = {1--7},
  year    = {2023},
  doi     = {10.1109/TASC.2023.3270766}
}

@article{karamuftuoglu2024unsupervised,
  author  = {Karamuftuoglu, Mustafa Altay and Ucpinar, Beyza Zeynep and Razmkhah, Sasan and Kamal, Mehdi and Pedram, Massoud},
  title   = {Unsupervised {SFQ}-Based Spiking Neural Network},
  journal = {IEEE Transactions on Applied Superconductivity},
  volume  = {34},
  number  = {3},
  pages   = {1--8},
  note    = {Art. no. 1300708},
  year    = {2024}
}

@article{karamuftuoglu2025ternary,
  author  = {Karamuftuoglu, Mustafa Altay and Ucpinar, Beyza Zeynep and Fayyazi, Arash and Razmkhah, Sasan and
             Kamal, Mehdi and Pedram, Massoud},
  title   = {Scalable Superconductor Neuron with Ternary Synaptic Connections for Ultra-Fast {SNN} Hardware},
  journal = {Superconductor Science and Technology},
  volume  = {38},
  number  = {2},
  pages   = {025014},
  year    = {2025},
  doi     = {10.1088/1361-6668/adaaa9}
}

@inproceedings{liu2023sushi,
  author    = {Liu, Zeshi and Chen, Shuo and Qu, Peiyao and Liu, Huanli and Niu, Minghui and Ying, Liliang and
               Ren, Jie and Tang, Guangming and You, Haihang},
  title     = {{SUSHI}: Ultra-High-Speed and Ultra-Low-Power Neuromorphic Chip Using Superconducting Single-Flux-Quantum Circuits},
  booktitle = {Proceedings of the 56th Annual IEEE/ACM International Symposium on Microarchitecture (MICRO)},
  pages     = {614--627},
  year      = {2023},
  doi       = {10.1145/3613424.3623787}
}

@article{onomi2008updown,
  author  = {Onomi, Takeshi and Kondo, Taizo and Nakajima, Koji},
  title   = {High-Speed Single Flux-Quantum Up/Down Counter for Neural Computation Using Stochastic Logic},
  journal = {Journal of Physics: Conference Series},
  volume  = {97},
  pages   = {012187},
  year    = {2008}
}

@article{han2026stochadder,
  author  = {Han, Zeyu and Mandai, Yuya and Yoshikawa, Nobuyuki and Yamanashi, Yuki},
  title   = {Small-Area and High-Precision Deterministic Stochastic Adder Using Single Flux Quantum Circuit},
  journal = {Superconductor Science and Technology},
  volume  = {39},
  number  = {1},
  pages   = {015003},
  year    = {2026},
  doi     = {10.1088/1361-6668/ae34ed}
}

@article{edwards2024stochastic,
  author  = {Edwards, Alexander J. and Krylov, Gleb and Friedman, Joseph S. and Friedman, Eby G.},
  title   = {Harnessing Stochasticity for Superconductive Multi-Layer Spike-Rate-Coded Neuromorphic Networks},
  journal = {Neuromorphic Computing and Engineering},
  volume  = {4},
  number  = {1},
  pages   = {014005},
  year    = {2024},
  doi     = {10.1088/2634-4386/ad207a}
}

\clearpage

\section*{Appendix A: Hard-Wired vs. Programmable Weights}
\label{app:weight_tradeoffs}

\subsection*{A.1 Rationale for the Hard-Wired Baseline}
The hardware results in the main text (Tables~VI and~VII) assume hard-wired weights. In this implementation, each select threshold $S_i$ is a constant, so the comparator $[R_k(t) < S_i]$ reduces to a fixed chain of AND/OR gates on the RNG bits, the sign routing is pure wiring, and no on-chip weight storage is needed (Sec.~IV-C). We adopt this baseline to isolate the savings of the stateless AND-NOT activation and the adder-free butterfly normal form from the cost of weight storage.

These savings do not depend on how the weights are stored. Removing the per-neuron ReLU counters, which account for 60--70\% of the JJs in our initial counter-based design (Sec.~III-E), and their feedback loops applies equally to hard-wired and programmable implementations.

\subsection*{A.2 Cost of Programmable Weights}
A deployment that runs several models on one chip requires programmable weights, e.g., NDRO cells or shift-register arrays that are loaded once per model (Sec.~III-A). Each select then needs a full comparator between the $B$-bit RNG value and the stored $(B{+}1)$-bit threshold, and the sign routing becomes a pair of small MUXes. For the decoder and the digits chain, this increases the logic by $6.8\times$ and $\approx 11\times$, to 3.4~M and 31~M JJs, respectively (Sec.~V-G). In addition, the weight registers hold 20~kbit (decoder), 183~kbit (digits chain), and up to 0.91~Mbit (flow chains), which at 12--16~JJs per NDRO bit adds about 0.24--0.32~M JJs for the decoder and 2.2--14.5~M JJs for the flow chains. For the Fashion-MNIST chain, the total programmable footprint is $\sim$75~M JJs.

The main conclusions of the paper still hold with programmable weights:

\begin{enumerate}
    \item \textbf{Area relative to a binary design:} Including logic and weight storage, the programmable Fashion-MNIST chain ($\sim$75~M JJs) is still about $4.3\times$ smaller than the bit-parallel fixed-point SFQ implementation of the same network (321~M JJs, Sec.~V-G).
    \item \textbf{Off-chip bandwidth:} The weights are loaded once per model and then read locally in every clock cycle, so they add no off-chip traffic during inference. As in the hard-wired design, the chip exchanges only input thresholds and output counts once per window, which requires 21~Gb/s. The bit-parallel binary design must exchange one sample per clock cycle with memory at a warmer temperature stage, which requires 28~Tb/s for the Fashion-MNIST chain (Sec.~V-G), i.e., about $1000\times$ more.
    \item \textbf{Time per sample:} The time per sample is set by the window length $T$ and the clock frequency, and the comparators lie off the critical path (Sec.~V-G). We therefore expect the programmable design to keep the same time per sample ($\approx 0.49~\mu$s at $T{=}4096$ and 9~GHz), with at most a slightly longer pipeline fill.
\end{enumerate}

The two implementations thus serve different purposes: hard-wired weights give the smallest fixed-function accelerator, whereas programmable weights allow one chip to run any model with the same architecture, at a 6.8--11$\times$ higher logic cost plus the weight storage.

\begin{strip}
\centering
\small
\captionof{table}{Comparison of hard-wired and programmable weight implementations ($B{=}8$, relaxed balancing, MIT-LL SFQ5ee cell library).}
\label{tab:hardwired_vs_programmable}
\begin{tabularx}{\textwidth}{@{}l X X@{}}
\toprule
\textbf{Metric} & \textbf{Hard-wired weights} & \textbf{Programmable weights} \\
\midrule
Flexibility & One fixed model & Any model with the same architecture \\
Select logic & Fixed AND/OR chain on RNG bits & Full comparator $[R_k(t) < S_i]$ \\
Sign routing & Wiring (channel swap) & Pair of small MUXes per cell \\
Weight storage & None & 20~kbit (decoder), 183~kbit (digits) to 0.91~Mbit (flow chains) \\
Storage JJs (12--16 JJs/bit) & 0 & 0.24--0.32~M (decoder); 2.2--14.5~M (flow chains) \\
\midrule
One stage, split ($n{=}64$, Table~VI) & 15.3--24.2~k JJs & 175~k JJs ($7.2$--$11.4\times$) \\
Saving from relaxed balancing & 40--47\% per stage & 26\% per stage \\
\midrule
VAE decoder ($n{=}64$) & 0.50~M JJs & 3.4~M JJs logic ($6.8\times$) + storage \\
Flow, digits ($n{=}128$) & 2.78~M JJs & 31~M JJs logic ($\approx 11\times$) + storage \\
Flow, Fashion-MNIST ($n{=}256$) & 6.2~M JJs & $\sim$75~M JJs total, incl.\ storage \\
\midrule
Weight loading & None (fixed at fabrication) & Once per model, negligible bandwidth \\
Off-chip bandwidth (Fashion) & 21~Gb/s & 21~Gb/s \\
Time per sample ($T{=}4096$, 9~GHz) & 0.46~$\mu$s (decoder), 0.49~$\mu$s (flows) & Expected unchanged \\
\bottomrule
\end{tabularx}
\end{strip}
\null
\clearpage

\end{document}